\documentclass{article}

\usepackage[nonatbib,preprint]{neurips_2018}

\usepackage{cite}
\usepackage[utf8]{inputenc} 
\usepackage[T1]{fontenc}    
\usepackage[hidelinks]{hyperref}       
\usepackage{url}            
\usepackage{booktabs}       
\usepackage{amsfonts}       
\usepackage{nicefrac}       
\usepackage{microtype}      
\usepackage{amsmath}
\usepackage{multirow}
\usepackage{graphicx}
\usepackage{xcolor}
\usepackage{epstopdf}
\usepackage{caption}
\usepackage{subcaption}
\usepackage{bm}
\usepackage{pgfplots}
\usepackage{footnote}
\usepackage{stfloats}
\usepackage{placeins}
\usepackage{color,soul}
\usepackage{dsfont}
\usepackage{amssymb}
\usepackage[ruled,vlined]{algorithm2e}
\usepackage{pbox}
\usepackage{enumerate}
\usepackage{etextools}
\usepackage{mathtools}
\usepackage{amsthm}
\usepackage{tabulary}
\usepackage{dsfont}
\usepackage{pgfplots}
\usepackage{bbm}
\pgfplotsset{compat=1.14}
\usepackage{comment}
\usepackage{algorithmicx}
\usepackage{subfiles}
\usepackage{wrapfig}

\newcommand{\rev}[1]{{\color{red} #1}}
\newcommand{\ejc}[1]{\textcolor{red}{[EJC: #1]}}

\newtheorem{thm}{Theorem} 
\newtheorem{lemma}{Lemma}

\newcommand{\ceil}[1]{\left\lceil#1\right\rceil}
\newcommand{\floor}[1]{\left\lfloor#1\right\rfloor}

\newcommand{\alphalo}{\alpha_{\text{\rm lo}}}
\newcommand{\alphahi}{\alpha_{\text{\rm hi}}}

\def\EE{{\mathbb E}}
\def\PP{{\mathbb P}}
\def\RR{{\mathbb R}}

\mathtoolsset{showonlyrefs}

\title{Conformalized Quantile Regression}

\author{%
  Yaniv Romano \\
  Department of Statistics \\
  Stanford University
  \And
  Evan Patterson \\
  Department of Statistics \\
  Stanford University
  \And
  Emmanuel J. Cand\`es \\
  Departments of Mathematics and of Statistics\\
  Stanford University
}

\begin{document}

\maketitle

\begin{abstract}

Conformal prediction is a technique for constructing prediction intervals that attain valid coverage in finite samples, without making distributional assumptions. Despite this appeal, existing conformal methods can be unnecessarily conservative because they form intervals of constant or weakly varying length across the input space. In this paper we propose a new method that is fully adaptive to heteroscedasticity. It combines conformal prediction with classical quantile regression, inheriting the advantages of both. We establish a theoretical guarantee of valid coverage, supplemented by extensive experiments on popular regression datasets. We compare the efficiency of conformalized quantile regression to other conformal methods, showing that our method tends to produce shorter intervals.

\end{abstract}

\section{Introduction} \label{sec:introduction}

In many applications of regression modeling, it is important not only to predict accurately but also to quantify the accuracy of the predictions. This is especially true in situations involving high-stakes decision making, such as estimating the efficacy of a drug or the risk of a credit default. The uncertainty in a prediction can be quantified using a prediction interval, giving lower and upper bounds between which the response variable lies with high probability. An ideal procedure for generating prediction intervals should satisfy two properties. First, it should provide valid coverage in finite samples, without making strong distributional assumptions, such as Gaussianity. Second, its intervals should be as short as possible at each point in the input space, so that the predictions will be informative. When the data is heteroscedastic, getting valid but short prediction intervals requires adjusting the lengths of the intervals according to the local variability at each query point in predictor space. This paper introduces a procedure that performs well on both criteria, being distribution-free and adaptive to heteroscedasticity.

Our work is heavily inspired by \emph{conformal prediction}, a general methodology for constructing prediction intervals \cite{vovk1999machine,papadopoulos2002inductive,vovk2005algorithmic, vovk2009line, lei2013distribution, lei2014distribution}. Conformal prediction has the virtue of providing a nonasymptotic, distribution-free coverage guarantee. The main idea is to fit a regression model on the training samples, then use the residuals on a held-out validation set to quantify the uncertainty in future predictions. The effect of the underlying model on the length of the prediction intervals, and attempts to construct intervals with locally varying length, have been studied in numerous recent works \cite{papadopoulos2008normalized, papadopoulos2008inductive,papadopoulos2011regression,johansson2014regression,johansson2014regression_local,lei2014distribution, johansson2015efficient, vovk2015cross, bostrom2017accelerating, lei2018distribution, chen2018discretized}. Nevertheless, existing methods yield conformal intervals of either fixed length or length depending only weakly on the predictors, as argued in \cite{lei2014distribution,lei2018distribution,vovk2019conformal}.

Quantile regression \cite{koenker1978regression} offers a different approach to constructing prediction intervals. Take any algorithm for \emph{quantile regression}, i.e., for estimating conditional quantile functions from data. To obtain prediction intervals with, say, nominal 90\% coverage, simply fit the conditional quantile function at the 5\% and 95\% levels and form the corresponding intervals. Even for highly heteroscedastic data, this methodology has been shown to be adaptive to local variability \cite{hunter2000quantile,taylor2000quantile,koenker2001quantile,meinshausen2006quantile,takeuchi2006nonparametric,steinwart2011estimating,tagasovska2018frequentist}. However, the validity of the estimated intervals is guaranteed only for specific models, under certain regularity and asymptotic conditions \cite{steinwart2011estimating,takeuchi2006nonparametric,meinshausen2006quantile}.

In this work, we combine conformal prediction with quantile regression. The resulting method, which we call \emph{conformalized quantile regression} (CQR), inherits both the finite sample, distribution-free validity of conformal prediction and the statistical efficiency of quantile regression.\footnote{Source code implementing CQR is available online at \url{https://github.com/yromano/cqr}.} On one hand, CQR is flexible in that it can wrap around any algorithm for quantile regression, including random forests and deep neural networks \cite{gal2016dropout,lian2016landslide,lakshminarayanan2017simple,pearce2018high}. On the other hand, a key strength of CQR is its rigorous control of the miscoverage rate, independent of the underlying regression algorithm.

\subsection*{Summary and outline} \label{sec:contribution}

Suppose we are given $ n $ training samples $ \{(X_i, Y_i)\}_{i=1}^n$ and we must now predict the unknown value of $Y_{n+1}$ at a test point $X_{n+1}$. We assume that all the samples $ \{(X_i,Y_i)\}_{i=1}^{n+1} $ are drawn exchangeably---for instance, they may be drawn i.i.d.---from an arbitrary joint distribution $P_{XY}$ over the feature vectors $ X\in \RR^p $ and response variables $ Y\in \RR $. We aim to construct a \emph{marginal distribution-free prediction interval} $C(X_{n+1}) \subseteq \RR$ that is likely to contain the unknown response $Y_{n+1} $. That is, given a desired miscoverage rate $ \alpha $, we ask that
\begin{align} \label{eq:cp_coverage}
\PP\{Y_{n+1} \in C(X_{n+1}) \} \geq 1-\alpha
\end{align}
for any joint distribution $ P_{XY} $ and any sample size $n$. The probability in this statement is marginal, being taken over all the samples $ \{(X_i, Y_i)\}_{i=1}^{n+1} $.

To accomplish this, we build on the method of conformal prediction \cite{papadopoulos2002inductive,papadopoulos2008inductive,vovk2005algorithmic}. We first split the training data into two disjoint subsets, a proper training set and a calibration set.\footnote{Like conformal regression, CQR has a variant that does not require data splitting.} We fit two quantile regressors on the proper training set to obtain initial estimates of the lower and upper bounds of the prediction interval, as explained in Section \ref{sec:q_reg}. Then, using the calibration set, we conformalize and, if necessary, correct this prediction interval. Unlike the original interval, the conformalized prediction interval is guaranteed to satisfy the coverage requirement \eqref{eq:cp_coverage} regardless of the choice or accuracy of the quantile regression estimator. We prove this in Section \ref{sec:conformal_qreg}.

Our method differs from the standard method of conformal prediction \cite{vovk2005algorithmic,lei2018distribution}, recalled in Section \ref{sec:conformal}, in that we calibrate the prediction interval using conditional quantile regression, while the standard method uses only classical, conditional mean regression. The result is that our intervals are adaptive to heteroscedasticity whereas the standard intervals are not. We evaluate the statistical efficiency of our framework by comparing its miscoverage rate and average interval length with those of other methods. We review existing state-of-the-art schemes for conformal prediction in Section \ref{sec:local_split_conformal} and we compare them with our method in Section \ref{sec:experiments}. Based on extensive experiments across eleven datasets, we conclude that conformal quantile regression yields shorter intervals than the competing methods.


\section{Quantile regression} \label{sec:q_reg}

The aim of conditional quantile regression \cite{koenker1978regression} is to estimate a given quantile, such as the median, of $ Y $ conditional on $ X $. Recall that the \emph{conditional distribution function} of $Y$ given $X=x$ is
\begin{align}
F(y \mid X=x) := \PP\{Y\leq y \mid X=x\},
\end{align}
and that the $ \alpha $th \emph{conditional quantile function} is
\begin{align} \label{eq:q}
q_{\alpha}(x) := \inf \{y \in \RR : F(y \mid X=x) \geq \alpha\}.
\end{align}

Fix the lower and upper quantiles to be equal to $ \alphalo = \alpha/2$ and $ \alphahi = 1 - \alpha/2 $, say. Given the pair $ q_{\alphalo}(x) $ and $ q_{\alphahi}(x) $ of lower and upper conditional quantile functions, we obtain a conditional prediction interval for $ Y $ given $ X=x $, with miscoverage rate $ \alpha $, as
\begin{align} \label{eq:q_interval}
C(x) = [q_{\alphalo}(x),\ q_{\alphahi}(x)].
\end{align}
By construction, this interval satisfies
\begin{align} \label{eq:q_coverage}
\PP\{Y \in C(X) | X=x \} \geq 1 - \alpha.
\end{align}
Notice that the length of the interval $C(X)$ can vary greatly depending on the value of $ X $. The uncertainty in the prediction of $ Y $ is naturally reflected in the length of the interval. In practice we cannot know this ideal prediction interval, but we can try to estimate it from the data.

\subsection*{Estimating quantiles from data}

Classical regression analysis estimates the conditional mean of the test response $ Y_{n+1} $ given the features $ X_{n+1}{=}x $ by minimizing the sum of squared residuals on the $n$ training points:
\begin{align} \label{eq:mu_optimization}
\hat{\mu}(x) = \mu(x; \hat\theta), \qquad \hat\theta = \underset{\theta}{\mathrm{argmin}} \ \frac{1}{n} \sum_{i=1}^n (Y_i - \mu(X_i ; \theta))^2 + \mathcal{R}(\theta).
\end{align}
Here $ \theta $ are the parameters of the regression model, $\mu(x; \theta)$ is the regression function, and $ \mathcal{R} $ is a potential regularizer. 

\begin{figure}[t]
	\centering
	\includegraphics[width=0.3\textwidth]{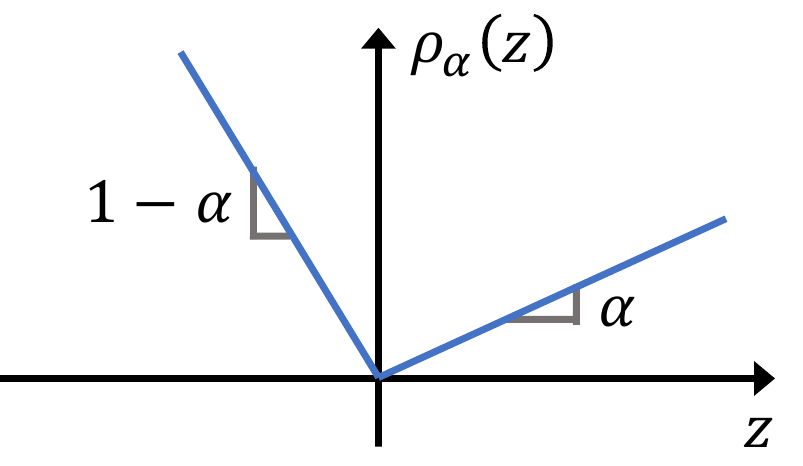}
	\caption{Visualization of the pinball loss function in \eqref{eq:pinball}, where $z=y-\hat{y}$.}
	\label{fig:pinball}
\end{figure}

Analogously, quantile regression estimates a conditional quantile function $q_{\alpha}$ of $ Y_{n+1} $ given $ X_{n+1} {=} x $. This can be cast as the optimization problem
\begin{align} \label{eq:q_optimization}
\hat{q}_\alpha(x) = f(x; \hat\theta), \qquad \hat\theta = \underset{\theta}{\mathrm{argmin}} \ \frac{1}{n} \sum_{i=1}^n \rho_\alpha (Y_i, f(X_i ; \theta))  + \mathcal{R}(\theta),
\end{align}
where $f(x;\theta)$ is the quantile regression function and the loss function $ \rho_\alpha $ is the ``check function'' or ``pinball loss'' \cite{koenker1978regression,steinwart2011estimating}, defined by
\begin{align} \label{eq:pinball}
\rho_\alpha(y, \hat{y}) := \begin{cases}
\alpha(y - \hat{y}) & \text{if } y - \hat{y}> 0, \\
(1-\alpha)(\hat{y} - y) & \text{otherwise}
\end{cases}
\end{align}
and illustrated in Figure \ref{fig:pinball}. 
The simplicity and generality of this formulation makes quantile regression widely applicable. As in classical regression, one can leverage the great variety of machine learning methods to design and learn $ \hat{q}_\alpha $ \cite{hunter2000quantile,taylor2000quantile,friedman2001greedy,koenker2001quantile,takeuchi2006nonparametric}.




All this suggests an obvious strategy to construct a prediction band with nominal miscoverage rate $\alpha$: estimate  $ \hat{q}_{\alphalo}(x) $ and $ \hat{q}_{\alphahi}(x) $ using quantile regression, then output $ \hat{C}(X_{n+1}) = [\hat{q}_{\alphalo}(X_{n+1}), \ \hat{q}_{\alphahi}(X_{n+1})] $ as an estimate of the ideal interval $C(X_{n+1})$ from equation \eqref{eq:q_interval}. This approach is widely applicable and often works well in practice, yielding intervals that are adaptive to heteroscedasticity. However, it is not guaranteed to satisfy the coverage statement \eqref{eq:q_coverage} when $ C(X) $ is replaced by the estimated interval $ \hat{C}(X_{n+1})$. Indeed, the absence of any finite sample guarantee can sometimes be disastrous. This worry is corroborated by our experiments, which show that the intervals constructed by neural networks can substantially undercover.

Under certain regularity conditions and for specific models, estimates of conditional quantile functions via the pinball loss are known to be asymptotically consistent \cite{steinwart2011estimating,takeuchi2006nonparametric}. Related methods that do not minimize the pinball loss, such as quantile random forests \cite{meinshausen2006quantile}, are also asymptotically consistent. But to get valid coverage in finite samples, we must draw on a different set of ideas, from conformal prediction.


\section{Conformal Prediction} \label{sec:conformal}

We now describe how conformal prediction \cite{vovk1999machine,vovk2005algorithmic} constructs prediction intervals that satisfy the finite-sample coverage guarantee \eqref{eq:cp_coverage}. To be carried out exactly, the original, or \emph{full}, conformal procedure effectively requires the regression algorithm to be invoked infinitely many times. In contrast, the method of \emph{split}, or \emph{inductive}, conformal prediction \cite{papadopoulos2002inductive, papadopoulos2008inductive} avoids this problem, at the cost of splitting the data. While our proposal is applicable to both versions of conformal prediction, in the interest of space we will restrict our attention to split conformal prediction and refer the reader to \cite{vovk2005algorithmic,lei2018distribution} for a more detailed comparison between the two methods.

Under the assumptions of Section \ref{sec:contribution}, the split conformal method begins by splitting the training data into two disjoint subsets: a proper training set $ \left\lbrace (X_i,Y_i): i \in \mathcal{I}_1 \right\rbrace  $ and calibration set $ \left\lbrace (X_i,Y_i): i \in \mathcal{I}_2 \right\rbrace  $. Then, given any regression algorithm $\mathcal{A}$,\footnote{In full conformal prediction, the regression algorithm must treat the data exchangeably, but no such restrictions apply to split conformal prediction.} a regression model is fit to the proper training set:
\begin{align}
\hat{\mu}(x) \leftarrow \mathcal{A} \left( \{(X_i,Y_i) : i \in \mathcal{I}_1\} \right).
\end{align}
Next, the absolute residuals are computed on the calibration set, as follows:
\begin{align} \label{eq:residual_err}
	R_i = |Y_i - \hat{\mu}(X_i)|, \qquad \ i \in \mathcal{I}_2.
\end{align}
For a given level $ \alpha $, we then compute a quantile of the empirical distribution\footnote{The explicit formula for empirical quantiles is recalled in Appendix \ref{app:lemmas}.} of the absolute residuals, 
\begin{align} \label{eq:quantilde_split}
Q_{1-\alpha}(R, \mathcal{I}_2) :=  (1-\alpha)(1+1/|\mathcal{I}_2|)\text{-th empirical quantile of} \left\{R_i : i \in \mathcal{I}_2 \right\}.
\end{align}
Finally, the prediction interval at a new point $ X_{n+1} $ is given by
\begin{align} \label{eq_c_cplit}
C(X_{n+1}) = \left[ \hat{\mu}(X_{n+1}) - Q_{1-\alpha}(R, \mathcal{I}_2) ,\ \hat{\mu}(X_{n+1}) + Q_{1-\alpha}(R, \mathcal{I}_2) \right].
\end{align}
This interval is guaranteed to satisfy \eqref{eq:cp_coverage}, as shown in  \cite{vovk2005algorithmic}. For related theoretical studies, see \cite{chen2016trimmed, lei2018distribution}. 

A closer look at the prediction interval \eqref{eq_c_cplit} reveals a major limitation of this procedure: the length of $ C(X_{n+1}) $ is fixed and equal to $ 2Q_{1-\alpha}(R, \mathcal{I}_2) $, independent of $ X_{n+1} $. Lei et al \cite{lei2018distribution} observe that the intervals produced by the full conformal method also vary only slightly with $ X_{n+1} $, provided the regression algorithm is moderately stable. This brings us to our proposal, which offers a principled approach to constructing variable-width conformal prediction intervals.

\section{Conformalized quantile regression (CQR)} \label{sec:conformal_qreg}

\begin{figure}[t]
	\centering
	\begin{subfigure}[a]{0.49\textwidth}
	\caption{Split: Avg. coverage 91.4\%; Avg. length 2.91.}
	\includegraphics[width=1\textwidth]{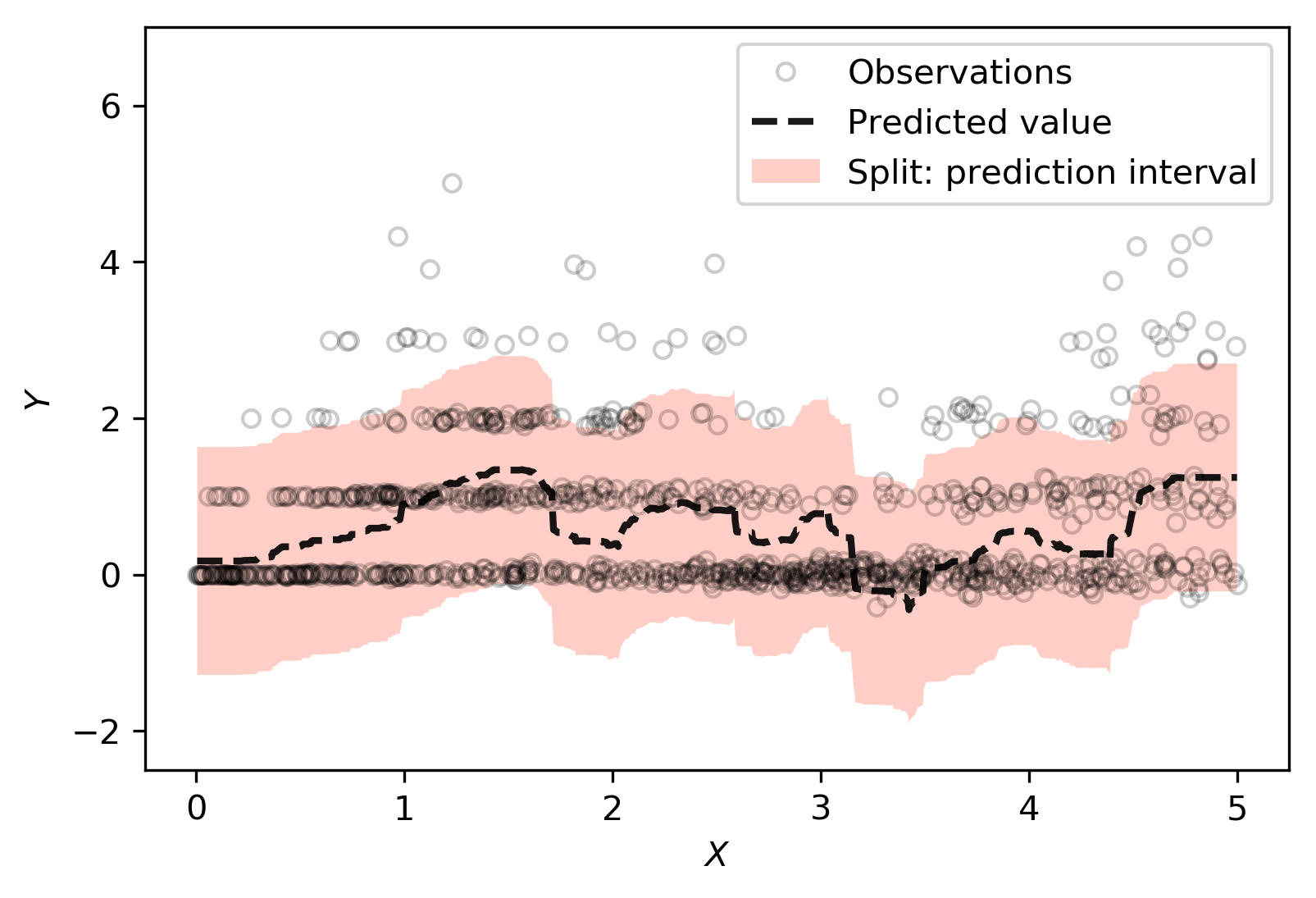}
	\label{subfig:split}
	\end{subfigure}
	\vspace{5pt}
	\begin{subfigure}[a]{0.49\textwidth}
	\caption{Local: Avg. coverage 91.7\%; Avg. length 2.86.}
	\includegraphics[width=1\textwidth]{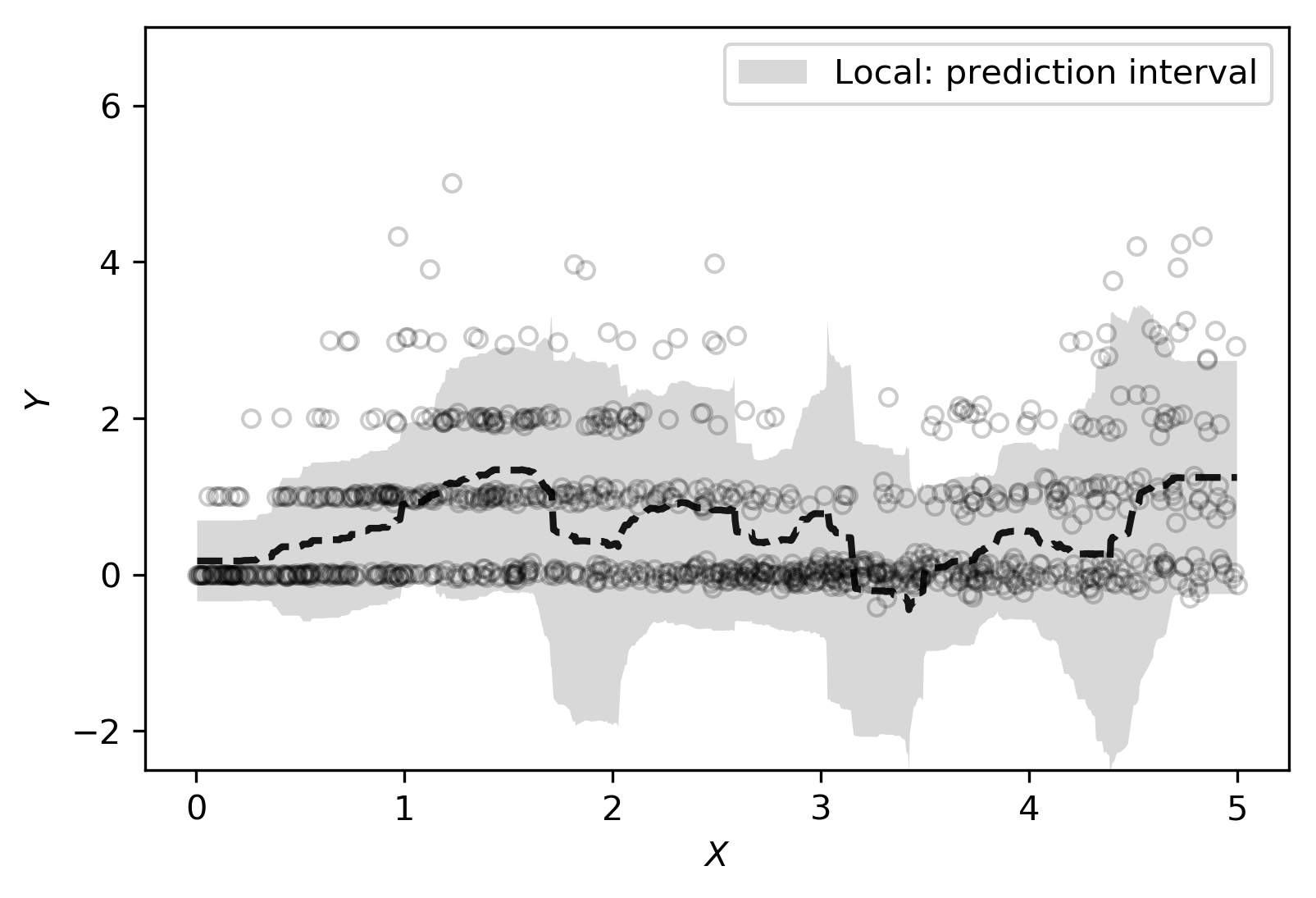}
	\label{subfig:local}
	\end{subfigure}
		\begin{subfigure}[a]{0.49\textwidth}
	\caption{CQR: Avg. coverage 91.06\%; Avg. length 1.99.}
	\includegraphics[width=1\textwidth]{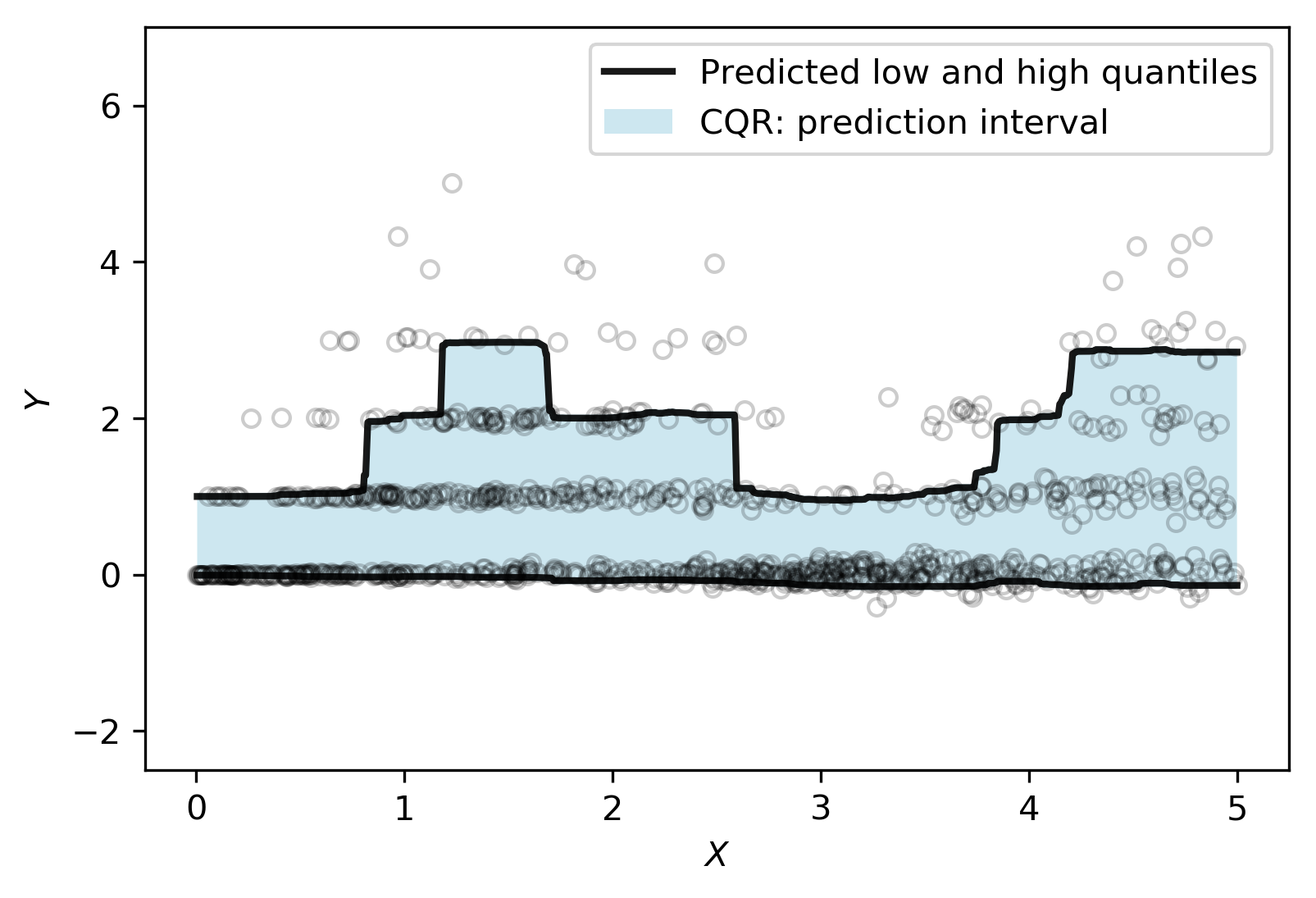}
	\label{subfig:cqr}
	\end{subfigure}
	\begin{subfigure}[a]{0.49\textwidth}
	\caption{Length of prediction intervals.}
	\includegraphics[width=1\textwidth]{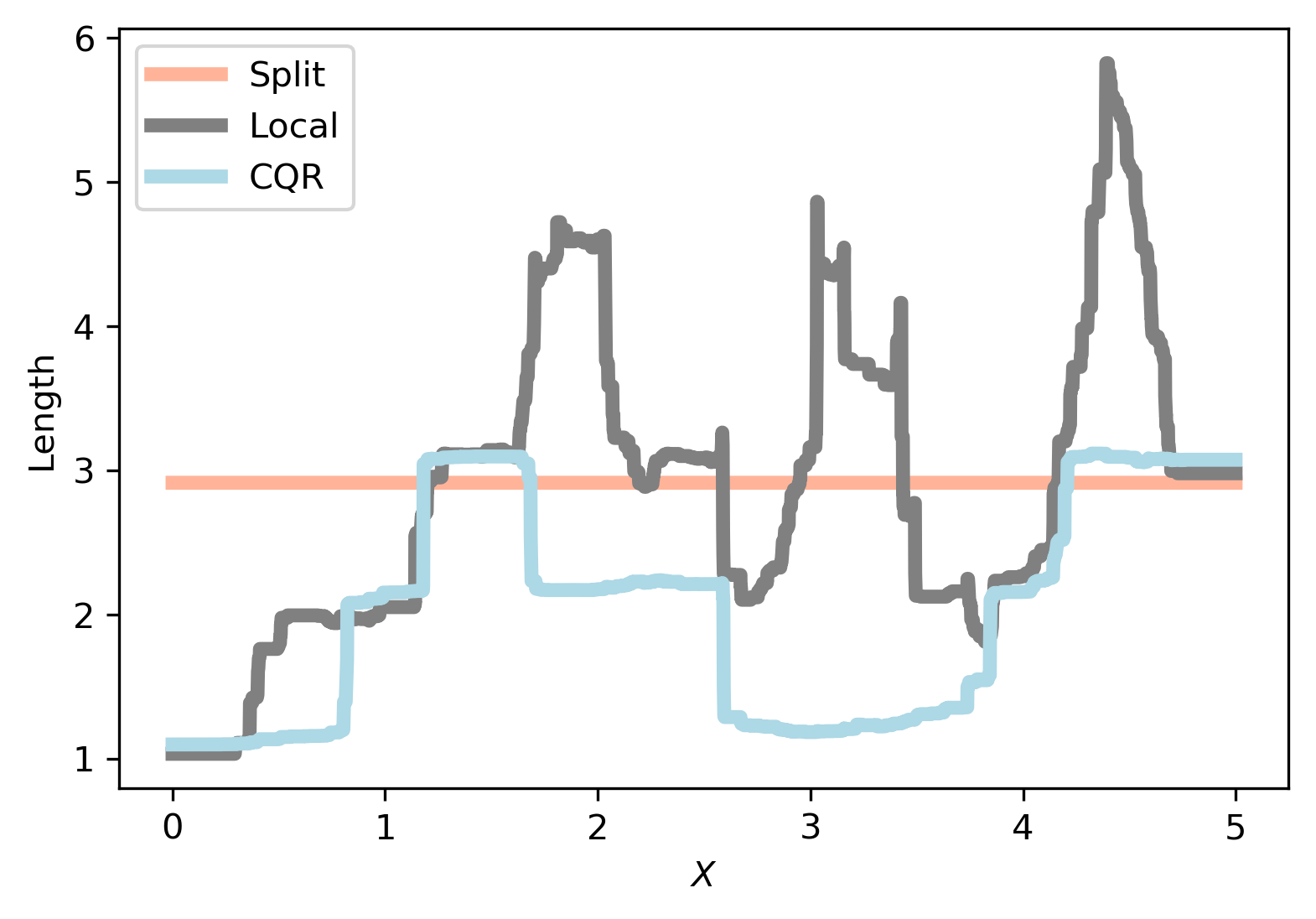}
	\label{subfig:length}
	\end{subfigure}
	\caption{Prediction intervals on simulated heteroscedastic data with outliers (see Figure \ref{fig:synthetic_illustration_full} for a full range display): (a) the standard split conformal method, (b) its locally adaptive variant, and (c) CQR (our method). The length of the interval as a function of $X$ is shown in (d). The target coverage rate is 90\%. The broken black curve in (a) and (b) is the pointwise prediction from the random forest estimator. In (c), we show two curves, representing the lower and upper quantile regression estimates based on random forests \cite{meinshausen2006quantile}. Observe how in this example the quantile regression estimates closely match the adjusted estimates---the boundary of the blue region---obtained by conformalization. }  
	\label{fig:synthetic_illustration}
\end{figure}

In this section we introduce our procedure, beginning with a small experiment on simulated data to show how it improves upon standard conformal prediction. Figure \ref{fig:synthetic_illustration} compares the prediction intervals produced by (a) the split conformal method, (b) its locally adaptive variant (described later in Section \ref{sec:local_split_conformal}), and (c) our method, conformalized quantile regression (CQR). The heteroskedasticity of the data is evident, as the dispersion of $Y$ varies considerably with $X$. The data also contains outliers, shown in Figure \ref{fig:synthetic_illustration_full} from Appendix \ref{app:synthetic_example}. For all three methods, we construct $90\%$ prediction intervals on the test data. From Figures \ref{subfig:split} and \ref{subfig:length}, we see that the lengths of the split conformal intervals are fixed and equal to $2.91$. The prediction intervals of the locally weighted variant, shown in Figure \ref{subfig:local}, are partially adaptive, resulting in slightly shorter intervals, of average length $2.86$. Our method, shown in Figure \ref{subfig:cqr}, is also adaptive, but its prediction intervals are considerably shorter, of average length $1.99$, due to better estimation of the lower and upper quantiles. We refer the reader to Appendix \ref{app:synthetic_example} for additional 
information about this experiment.

We now describe CQR itself. As in split conformal prediction, we begin by splitting the data into a proper training set, indexed by $\mathcal{I}_1$, and a calibration set, indexed by $\mathcal{I}_2$. Given any quantile regression algorithm $\mathcal{A}$, we then fit two conditional quantile functions $\hat{q}_{\alphalo}$ and $\hat{q}_{\alphahi}$ on the proper training set:
\begin{align}
	\left\lbrace \hat{q}_{\alphalo}, \hat{q}_{\alphahi} \right\rbrace \leftarrow \mathcal{A}(\left\lbrace (X_i, Y_i): i \in \mathcal{I}_1 \right\rbrace).
\end{align}
In the essential next step, we compute \emph{conformity scores} that quantify the error made by the plug-in prediction interval $ \hat{C}(x) = [\hat{q}_{\alphalo}(x), \ \hat{q}_{\alphahi}(x)]  $. The scores are evaluated on the calibration set as
\begin{align} \label{eq:qreg_conformity}
	E_i := \max\{\hat{q}_{\alphalo}(X_i) - Y_i, Y_i - \hat{q}_{\alphahi}(X_i)\}
\end{align}
for each $i \in \mathcal{I}_2$. The conformity score $E_i$ has the following interpretation. If $Y_i$ is below the lower endpoint of the interval, $Y_i < \hat{q}_{\alphalo}(X_i)$, then $E_i = |Y_i - \hat{q}_{\alphalo}(X_i)|$ is the magnitude of the error incurred by this mistake. Similarly, if $Y_i$ is above the upper endpoint of the interval, $Y_i > \hat{q}_{\alphahi}(X_i)$, then $E_i = |Y_i - \hat{q}_{\alphahi}(X_i)|$. Finally, if $Y_i$ correctly belongs to the interval, $\hat{q}_{\alphalo}(X_i) \leq Y_i \leq \hat{q}_{\alphahi}(X_i)$, then $E_i$ is the larger of the two non-positive numbers $\hat{q}_{\alphalo}(X_i) - Y_i$ and $Y_i - \hat{q}_{\alphahi}(X_i)$ and so is itself non-positive. The conformity score thus accounts for both undercoverage and overcoverage.

Finally, given new input data $X_{n+1}$, we construct the prediction interval for $Y_{n+1}$ as
\begin{align} \label{eq:c_cplit_qreg}
C(X_{n+1}) = \left[ \hat{q}_{\alphalo}(X_{n+1}) - Q_{1-\alpha}(E, \mathcal{I}_2) , \ \hat{q}_{\alphahi}(X_{n+1}) + Q_{1-\alpha}(E, \mathcal{I}_2) \right],
\end{align}
where 
\begin{align} \label{eq:qreg_quantilde_split}
Q_{1-\alpha}(E, \mathcal{I}_2) :=  (1-\alpha)(1+1/|\mathcal{I}_2|)\text{-th empirical quantile of} \left\{E_i : i \in \mathcal{I}_2\right\}
\end{align}
conformalizes the plug-in prediction interval.

For ease of reference, the CQR procedure is summarized in Algorithm \ref{alg:split_qreg}. We now prove that its prediction intervals satisfy the marginal, distribution-free coverage guarantee \eqref{eq:cp_coverage}.

\begin{algorithm}[t]
	\caption{Split Conformal Quantile Regression.}
	\label{alg:split_qreg}
	
	\textbf{Input:}
	\begin{algorithmic}
		\State Data $(X_i, Y_i) \in \RR^p \times \RR, \ 1\leq i \leq n$.
		\State Miscoverage level $\alpha \in (0,1)$.
		\State Quantile regression algorithm $ \mathcal{A} $.
	\end{algorithmic}
	
	\textbf{Process:}
	\begin{algorithmic}
		\State Randomly split $ \left\lbrace 1,\dots,n \right\rbrace  $ into two disjoint sets $ \mathcal{I}_1 $ and $ \mathcal{I}_2 $.
		\State Fit two conditional quantile functions:
		$ \left\lbrace  \hat{q}_{\alphalo}, \hat{q}_{\alphahi} \right\rbrace \leftarrow \mathcal{A}(\left\lbrace (X_i, Y_i): i \in \mathcal{I}_1 \right\rbrace) $.
		\State Compute $ E_i $ for each $i \in \mathcal{I}_2$, as in equation \eqref{eq:qreg_conformity}.
		\State Compute $Q_{1-\alpha}(E, \mathcal{I}_2)$, the $(1-\alpha)(1+1/|\mathcal{I}_2|)$-th empirical quantile  of $\left\{E_i : i \in \mathcal{I}_2 \right\}$.
	\end{algorithmic}
	
	\textbf{Output:}
	\begin{algorithmic}
		\State Prediction interval $ C(x) = \left[ \hat{q}_{\alphalo}(x) - Q_{1-\alpha}(E, \mathcal{I}_2) , \ \hat{q}_{\alphahi}(x) + Q_{1-\alpha}(E, \mathcal{I}_2) \right] $ for unseen input $ X_{n+1}={x} $.
	\end{algorithmic}
	
\end{algorithm}

\begin{thm} \label{thm:validity_qreg}
	If $ (X_i, Y_i) $, $ i=1,\dots, n+1 $ are exchangeable, then the prediction interval $ C(X_{n+1}) $ constructed by the  split CQR algorithm satisfies 
	\begin{equation}
	\PP\{Y_{n+1} \in C(X_{n+1}) \} \geq 1-\alpha.
	\end{equation}
    Moreover, if the conformity scores $E_i$ are almost surely distinct, then the prediction interval is nearly perfectly calibrated:
    \begin{equation}
    \PP\{Y_{n+1} \in C(X_{n+1})\} \leq 1-\alpha+\frac{1}{|\mathcal{I}_2|+1}.
    \end{equation}
\end{thm}
\begin{proof}
	The result even holds, and we will prove it, conditionally on the proper training set.
	
	Let $E_{n+1}$ be the conformity score \eqref{eq:qreg_conformity} at the test point $(X_{n+1},Y_{n+1})$. By the construction of the prediction interval, we have
	\begin{align}
	    Y_{n+1} \in C(X_{n+1})
	    \quad\text{if and only if}\quad
	    E_{n+1} \leq Q_{1-\alpha}(E, \mathcal{I}_2),
	\end{align}
	and, in particular,
	\begin{equation} \label{eq:conditional-coverage}
	    \PP\{Y_{n+1} \in C(X_{n+1}) \,|\, (X_i,Y_i): i \in \mathcal{I}_1\} =
	    \PP\{E_{n+1} \leq Q_{1-\alpha}(E,\mathcal{I}_2)  \,|\, (X_i,Y_i): i \in \mathcal{I}_1\}.
	\end{equation}
	Since the original pairs $(X_i,Y_i)$ are exchangeable, so are the calibration variables $E_i$ for $i \in \mathcal{I}_2$ and $i=n+1$. Therefore, by Lemma \ref{lemma:quantile-inflation} on inflated empirical quantiles (stated in Appendix \ref{app:lemmas}),
	\begin{equation} \label{eq:conditional-quantiles-lower}
	    \PP\{E_{n+1} \leq Q_{1-\alpha}(E,\mathcal{I}_2)  \,|\, (X_i,Y_i): i \in \mathcal{I}_1\}
	    \geq 1 - \alpha,
	\end{equation}
	and, under the additional assumption that the $E_i$'s are almost surely distinct,
	\begin{equation} \label{eq:conditional-quantiles-upper}
	    \PP\{E_{n+1} \leq Q_{1-\alpha}(E,\mathcal{I}_2)  \,|\, (X_i,Y_i): i \in \mathcal{I}_1\}
	    \leq 1 - \alpha + \frac{1}{|\mathcal{I}_2|+1}.
	\end{equation}
    The result follows by taking expectations over the proper training set in \eqref{eq:conditional-coverage}, \eqref{eq:conditional-quantiles-lower}, and \eqref{eq:conditional-quantiles-upper}.
\end{proof}

\subsection*{Practical considerations and extensions} \label{sec:impl_details}

Conformalized quantile regression can accommodate a wide range of quantile regression methods \cite{koenker1978regression,hunter2000quantile,taylor2000quantile,friedman2001greedy,koenker2001quantile,meinshausen2006quantile,takeuchi2006nonparametric,tagasovska2018frequentist} to estimate the conditional quantile functions, $ q_{\alphalo} $ and $ q_{\alphahi} $. The estimators can be even be aggregates of different quantile regression algorithms. Recently, new deep learning techniques have been proposed \cite{gal2016dropout,lian2016landslide,lakshminarayanan2017simple,pearce2018high} for constructing prediction intervals. These methods could be wrapped by our framework and would then immediately enjoy rigorous coverage guarantees. In our experiments, we focus on quantile neural networks \cite{taylor2000quantile} and quantile regression forests \cite{meinshausen2006quantile}.

Because the underlying quantile regression algorithm may process the proper training set in arbitrary ways, our framework affords broad flexibility in hyper-parameter tuning. Consider, for instance, the tuning of typical hyper-parameters of neural networks, such as the batch size, the learning rate, and the number of epochs. The hyperparameters may be selected, as usual, by cross validation, where we minimize the average interval length over the folds.

In this vein, we record two specific implementation details that we have found to be useful. 
\begin{enumerate}
\item Quantile regression is sometimes \emph{too} conservative, resulting in unnecessarily wide prediction intervals. In our experience, quantile regression forests \cite{meinshausen2006quantile} are often overly conservative and quantile neural networks \cite{taylor2000quantile} are occasionally so. We can mitigate this problem by tuning the nominal quantiles of the underlying method as additional hyper-parameters in cross validation. Notably, this tuning does not invalidate the coverage guarantee, but it may yield shorter intervals, as our experiments confirm.

\item To reduce the computational cost, instead of fitting two separate neural networks to estimate the lower and upper quantile functions, we can replace the standard one-dimensional estimate of the unknown response by a two-dimensional estimate of the lower and upper quantiles. In this way, most of the network parameters are shared between the two quantile estimators. We adopt this approach in the experiments of Section \ref{sec:experiments}.
\end{enumerate}

Another avenue for extension is the conformalization step. The conformalization implemented by equations \eqref{eq:c_cplit_qreg} and \eqref{eq:qreg_quantilde_split} allows coverage errors to be spread arbitrarily over the left and right tails.
By using a method reminiscent of  \cite{linusson2014signed}, we can control the left and right tails independently, resulting in a stronger coverage guarantee.

\begin{thm} \label{thm:validity_qreg_asymmetric}
    Define the prediction interval
    \begin{equation}
    C(X_{n+1}) := [\hat{q}_{\alphalo}(X_{n+1}) - Q_{1-\alphalo}(E_{\mathrm{lo}}, \mathcal{I}_2),\; \hat{q}_{\alphahi}(X_{n+1}) + Q_{1-\alphahi}(E_{\mathrm{hi}}, \mathcal{I}_2)],
    \end{equation}
    where $Q_{1-\alphalo}(E_{\mathrm{lo}}, \mathcal{I}_2)$ is the $(1-\alphalo)$-th empirical quantile of $\{\hat{q}_{\alphalo}(X_i) - Y_i: i \in \mathcal{I}_2\}$ and $Q_{1-\alphahi}(E_{\mathrm{hi}}, \mathcal{I}_2)$ is the $(1-\alphahi)$-th empirical quantile of $\{Y_i - \hat{q}_{\alphahi}(X_i): i \in \mathcal{I}_2\}$. If the samples $ (X_i, Y_i) $, $ i=1,\dots, n+1 $ are exchangeable, then
    \begin{equation} \label{eq:asymmetric-coverage-lower}
        \PP\{Y_{n+1} \geq \hat{q}_{\alphalo}(X_{n+1}) - Q_{1-\alphalo}(E_{\mathrm{lo}}, \mathcal{I}_2)\}
        \geq 1-\alphalo
    \end{equation}
    and
    \begin{equation} \label{eq:asymmetric-coverage-upper}
        \PP\{Y_{n+1} \leq \hat{q}_{\alphahi}(X_{n+1}) + Q_{1-\alphahi}(E_{\mathrm{hi}}, \mathcal{I}_2)\}
        \geq 1-\alphahi.
    \end{equation}
    Consequently, assuming $\alpha = \alphalo + \alphahi$, we also have $\PP\{Y_{n+1} \in C(X_{n+1})\} \geq 1 - \alpha$.
\end{thm}
\begin{proof}
The two events inside the probabilities \eqref{eq:asymmetric-coverage-lower} and \eqref{eq:asymmetric-coverage-upper} are equivalent to $\hat{q}_{\alphalo}(X_{n+1}) - Y_{n+1} \leq Q_{1-\alphalo}(E_{\mathrm{lo}},\mathcal{I}_2)$ and $Y_{n+1} - \hat{q}_{\alphahi}(X_{n+1}) \leq Q_{1-\alphahi}(E_{\mathrm{hi}},\mathcal{I}_2)$, respectively. We can thus apply \mbox{Lemma \ref{lemma:quantile-inflation}} twice, in the same manner as in the proof of Theorem \ref{thm:validity_qreg}.
\end{proof}

As we will see in Section \ref{sec:experiments}, the price paid for the stronger coverage guarantee is slightly longer intervals.

\section{Related work: locally adaptive  conformal prediction} \label{sec:local_split_conformal}

\emph{Locally adaptive split conformal prediction}, first proposed in \cite{papadopoulos2008normalized,papadopoulos2011regression} and later studied in \cite{lei2018distribution}, is an earlier approach to making conformal prediction adaptive to heteroskedascity. Like our method, it starts from the observation that one can replace the absolute residuals in equation \eqref{eq:residual_err} by any other loss function that treats the data exchangeably. In this case, the absolute residuals $R_i$ are replaced by the scaled residuals 
\begin{align} \label{eq:locally_adaptive_residuals}
\tilde{R}_i := \frac{|Y_i - \hat{\mu}(X_i)|}{\hat{\sigma}(X_i)} = \frac{R_i}{\hat{\sigma}(X_i)}, \qquad \ i\in \mathcal{I}_2,
\end{align}
where $ \hat{\sigma}(X_i) $ is a measure of the dispersion of the residuals at $X_i$. Usually $ \hat{\sigma}(x) $ is an estimate of the conditional mean absolute deviation (MAD) of $ | Y - \hat\mu(x) | $ given $ X=x $. Finally, the prediction interval at a new point $ X_{n+1}$ is computed as
\begin{align} \label{eq:c_cplit_adaptive}
C(X_{n+1}) = \left[ \hat{\mu}(X_{n+1}) - \hat{\sigma}(X_{n+1}) Q_{1-\alpha}(\tilde{R}, \mathcal{I}_2) , \ \hat{\mu}(X_{n+1}) + \hat{\sigma}(X_{n+1}) Q_{1-\alpha}(\tilde{R}, \mathcal{I}_2) \right].
\end{align}
Both $ \hat{\mu} $ and $ \hat{\sigma} $ are fit only on the proper training set. Consequently, $ \hat{\mu} $ and $ \hat{\sigma} $ satisfy the assumptions of conformal prediction and, hence, locally adaptive conformal prediction inherits the coverage guarantee of standard conformal prediction.

In practice, locally adaptive conformal prediction requires fitting two functions, in sequence, on the proper training set. (Thus it is more computationally expensive than standard conformal prediction.) First, one fits the conditional mean function $ \hat{\mu}(x) $, as described in Section \ref{sec:conformal}. Then one fits $ \hat{\sigma}(x) $ to the pairs $  \left\lbrace (X_i, R_i): i\in \mathcal{I}_1 \right\rbrace $, 
using a regression model that predicts the residuals $ R_i $ given the inputs $ X_i $. As an example, the intervals in Figure \ref{subfig:local} above are created by locally adaptive split conformal prediction, where both $\hat{\mu}$ and $\hat{\sigma}$ are random forests.

Practitioners employ various tweaks to improve the method's numerical stability and statistical performance. Following \cite{johansson2014regression}, we add a hyper-parameter $\gamma > 0$ as a constant offset to the scale estimator $\hat\sigma(x)$. The scaled residuals then become
\begin{align} \label{eq:locally_adaptive_residuals_offset}
\tilde{R}_i = \frac{R_i}{\hat{\sigma}(X_i) + \gamma}.
\end{align}

\subsection*{Limitations of locally adaptive conformal prediction}

Locally adaptive conformal prediction is limited in several ways, some more important than others. 
A first limitation, already noted in \cite{lei2018distribution}, appears when the data is actually homoskedastic. In this case, the locally adaptive method suffers from inflated prediction intervals compared to the standard method. This is presumably due to the extra variability introduced by estimating $ \hat{\sigma} $ as well as $ \hat\mu $.

The locally adaptive method faces a more fundamental statistical limitation. There is an essential difference between the residuals on the proper training set and the residuals on the calibration set: the former are biased by an optimization procedure designed to minimize them, while the latter are unbiased. Because it uses the proper training residuals (as it must to ensure valid coverage), the locally adaptive method tends to systematically underestimate the prediction error. In general, this forces the correction constant $ Q_{1-\alpha}(\tilde{R}, \mathcal{I}_2) $ to be large and the intervals to be less adaptive than they could be.

To press this point further, suppose the conditional mean function $\hat\mu$ is a deep neural network. It is well attested in the deep learning literature that, given enough training samples, the best prediction error is attained by ``over-fitting'' to the training data, in the sense that the training error is nearly zero. The training residuals are then very poor estimates of the test residuals, resulting in severe loss of adaptivity. The original training objective of our method, in contrast, is to estimate the lower and upper conditional quantiles, not the conditional mean. Having sufficient training data, the fitted network is expected to provide reasonable approximations of these two quantile functions, which are used to construct adaptive prediction intervals.

\section{Experiments} \label{sec:experiments}

In this section we systematically compare our method, conformalized quantile regression, to the standard and locally adaptive versions of split conformal prediction. Among preexisting conformal prediction algorithms, we select leading variants that use random forests \cite{johansson2014regression} and neural networks \cite{papadopoulos2011reliable} for conditional mean regression. Likewise, we configure our method to use quantile regression algorithms based on random forests \cite{meinshausen2006quantile} and neural networks \cite{taylor2000quantile}. As a baseline, we also include conformal ridge regression \cite{papadopoulos2002inductive} in the comparison. A detailed description of each of the methods is given below.

We conduct the experiments on eleven popular benchmark datasets for regression, listed in Section \ref{sec:detatiled_experiments}. In each case, we standardize the features to have zero mean and unit variance and we rescale the response by dividing it by its mean absolute value.\footnote{In the experiments, we compute the needed sample means and variances only on the proper training set. This ensures that if the original data is exchangeable, then the rescaled data remains so. That being said, we could also rescale using sample means and variances computed on the test data, because it would preserve exchangeability even while it destroys independence.} The performance metrics are averaged over $ 20 $ different training-test splits; $ 80\% $ of the examples are used for training and the remaining $ 20\% $ for testing. The proper training and calibration sets for split conformal prediction have equal size. Throughout the experiments the nominal miscoverage rate is fixed and equal to $ \alpha=0.1 $.

\subsection{Methods} \label{sec:experiments_methods}

In more detail, we compare the following methods related to conformal prediction. We evaluate the original version of split conformal prediction (Section \ref{sec:conformal}) using the following three regression algorithms.
 
\begin{itemize}
	\item \textbf{Ridge}: We include ridge regression as a baseline. The regularization parameter is tuned by cross validation.
	
	\item \textbf{Random Forests}: We use the implementation of (conditional mean) random forest regression in the Python package {\tt sklearn}. The hyper-parameters are the package defaults, except for the total number of trees in the forest, which we set to $1000$.
	
	\item \textbf{Neural Net}: Our neural network architecture consists of three fully connected layers, with ReLU nonlinearities between layers. The first layer takes as input the $ p $-dimensional feature vector $ X $ and outputs $ 64 $ hidden variables. The second layer follows the same template, outputting another $ 64 $ hidden variables. Finally, a linear output layer returns a pointwise estimate of the response variable $ Y $. The parameters of the network are fit by minimizing the quadratic loss function \eqref{eq:mu_optimization}. We use the stochastic optimization algorithm Adam \cite{kingma2014adam}, with fixed learning rate of $ 5 \times 10^{-4} $, minibatches of size $ 64 $, and weight decay parameter equal to $ 10^{-6} $. We employ dropout regularization \cite{srivastava2014dropout}, with the probability of retaining a hidden unit equal to $0.1$. To avoid overfitting, we found that early stopping  performs well; we tune the number of epochs by cross validation, with an upper limit of $1000$ epochs.
\end{itemize}

We evaluate locally adaptive conformal prediction (Section \ref{sec:local_split_conformal}) using the same three underlying regression algorithms. We set the hyper-parameter $\gamma$ in equation \eqref{eq:locally_adaptive_residuals_offset} to $1$, which improves performance considerably compared to $\gamma = 0$.

\begin{itemize}
	\item \textbf{Ridge Local}: The conditional mean estimator $ \hat{\mu} $ is fit by ridge regression, as described above, and the mean absolute deviation (MAD) estimator $ \hat{\sigma} $ is $ k $-nearest neighbors with $ k=11 $.
	
	\item \textbf{Random Forests Local}: Both $ \hat{\mu} $ and $ \hat{\sigma} $ are random forests with the hyper-parameters described above.
	
	\item \textbf{Neural Net Local}: Both $ \hat{\mu} $ and $ \hat{\sigma} $ are neural networks, with the network architecture, hyper-parameters, and training algorithm described above.
\end{itemize}

For our own proposal, conformalized quantile regression (Algorithm \ref{alg:split_qreg}), we evaluate two variants:

\begin{itemize}
	\item \textbf{CQR Random Forests}: We use CQR with quantile regression forests \cite{meinshausen2006quantile}. To ensure a fair comparison, the hyper-parameters of the quantile regression forests are made identical to those of the random forests in the previous methods. Quantile regression forests have two additional parameters that control the coverage rate on the training data. We tune them using cross validation, as explained in Section \ref{sec:impl_details}.
	
	\item \textbf{CQR Neural Net}: We apply CQR using neural networks for quantile regression \cite{taylor2000quantile}. The network architecture is the same as above, except that the output of the quantile regression network is a two-dimensional vector, representing the lower and upper conditional quantiles. The training algorithm is also the same, except that the cost function is now the pinball loss in equation \eqref{eq:q_optimization} instead of the quadratic loss.
\end{itemize}

Finally, for the sake of comparison, we also include the previous two quantile regression algorithms, but without any conformalization:

\begin{itemize}
	\item \textbf{Quantile Random Forests}: We use quantile regression forests with hyperparameters as in the CQR procedure, except that the upper and lower levels are fixed at $0.05$ and $0.95$.
	
	\item \textbf{Quantile Neural Net}: We use quantile regression neural networks with exactly the same architecture and training algorithm as in the CQR procedure.
\end{itemize}

Unlike the preceding methods, the last two methods do not need a calibration set and do not have a finite-sample coverage guarantee. We fit them on the entire training set.

\subsection{Summary of results}

\begin{table}[]
\centering
\begin{tabular}{@{}lcc@{}}
\toprule
\textbf{Method}          & \textbf{Avg. Length} & \textbf{Avg. Coverage} \\ \midrule
Ridge                    & 3.06                 & 90.03                  \\
Ridge Local              & 2.94                 & 90.13                  \\
Random Forests           & 2.24                 & 89.99                  \\
Random Forests Local     & 1.82                 & 89.95                  \\
Neural Net               & 2.16                 & 89.92                  \\
Neural Net Local         & 1.81                 & 89.95                  \\
\textbf{CQR Random Forests}       & \textbf{1.41}                 & \textbf{90.33}                  \\
\textbf{CQR Neural Net}           & \textbf{1.40}                 & \textbf{90.05}                  \\
*Quantile Random Forests & *2.23                 & *92.62                  \\
*Quantile Neural Net     & *1.49                 & *88.51                  \\ \bottomrule
\end{tabular}
\vspace{5pt}
\caption{Length and coverage of prediction intervals ($ \alpha =0.1$) constructed by various methods, averaged across 11 datasets and 20 random training-test splits. Our methods are shown in bold font. The methods marked by an asterisk are not supported by finite-sample coverage guarantees.}
\label{fig:results}
\end{table}

Table \ref{fig:results} summarizes our 2,200 experiments, showing the average performance across all the datasets and training-test splits. On average, our method achieves shorter prediction intervals than both standard and locally adaptive conformal prediction. It may seem surprising that our method also outperforms non-conformalized quantile regression, which is permitted more training data. There are several possible explanations for this. First, the non-conformalized methods sometimes  \emph{over}cover, but that is mitigated by our signed conformity scores \eqref{eq:qreg_conformity}. In addition, by using CQR, we can tune the quantiles of the underlying quantile regression algorithms using cross-validation (Section \ref{sec:impl_details}). Interestingly, CQR selects quantiles below the nominal level.

Turning to the issue of valid coverage, all methods based on conformal prediction successfully construct prediction bands at the nominal coverage rate of $90\%$, as the theory suggests they should. One of the non-conformalized methods, based on random forests, is slightly conservative, while the other, based on neural networks, tends to undercover. In fact, other authors have shown that the coverage of quantile neural networks depends greatly on the tuning of the hyper-parameters, with, for instance, the actual coverage in \cite[Figure 3]{tagasovska2018frequentist} ranging from the 95\% nominal level in that paper to well below 50\%. Such volatility demonstrates the importance of the conformal prediction's finite-sample guarantee.

When estimating a lower and an upper quantile by two separate quantile regressions, there is no guarantee that the lower estimate will actually be smaller than the upper estimate. This is known as the \emph{quantile crossing} problem \cite{bassett1982empirical}. Quantile crossing can affect quantile neural networks, but not quantile regression forests. When the two quantiles are far apart, as in the 5\% and 95\% quantiles, we should expect the estimates to cross very infrequently and that is indeed what we find in the experiments. Nevertheless, we also evaluated a post-processing method to eliminate crossings \cite{chernozhukov2010quantile}. It yields a slight improvement in performance: the average interval length of the CQR neural networks drops from 1.40 to 1.35 and the average interval length of the unconformalized quantile neural networks drops from 1.49 to 1.41, with the coverage rates remaining about the same.

As expected, adopting the two-tailed, asymmetric conformalization proposed in Theorem \ref{thm:validity_qreg_asymmetric} causes an increase in average interval length compared to the symmetric conformalization of Theorem \ref{thm:validity_qreg}. Specifically, the average length for CQR neural networks increases from 1.40 to 1.58, while the coverage rate stays about the same. The average length for the CQR random forests increases from 1.41 to 1.57, accompanied by a slight increase in the average coverage rate, from 90.33 to 90.99.

\subsection{Performance on individual datasets} \label{sec:detatiled_experiments}

\begin{figure}[th]
	\centering
	\includegraphics[width=0.95\textwidth]{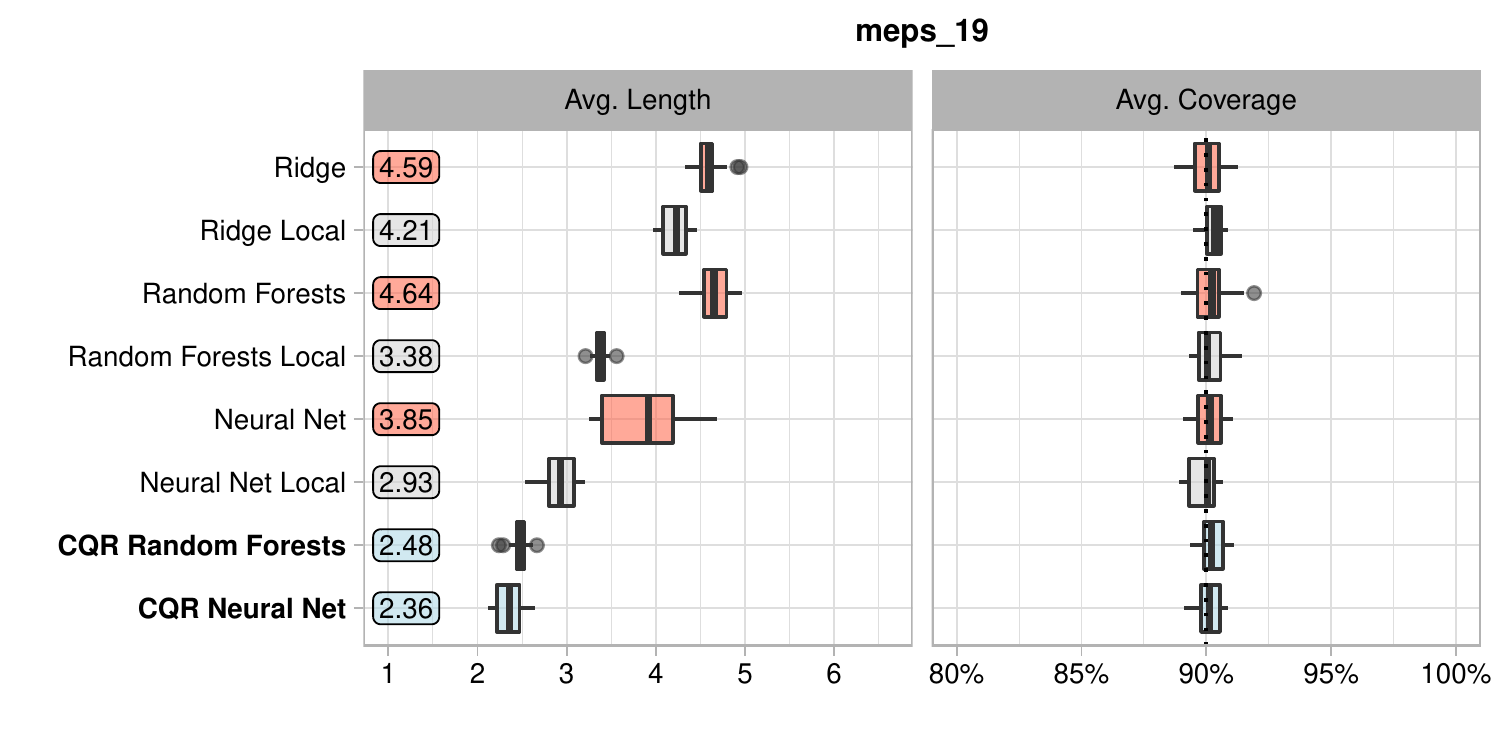}
	\includegraphics[width=0.95\textwidth]{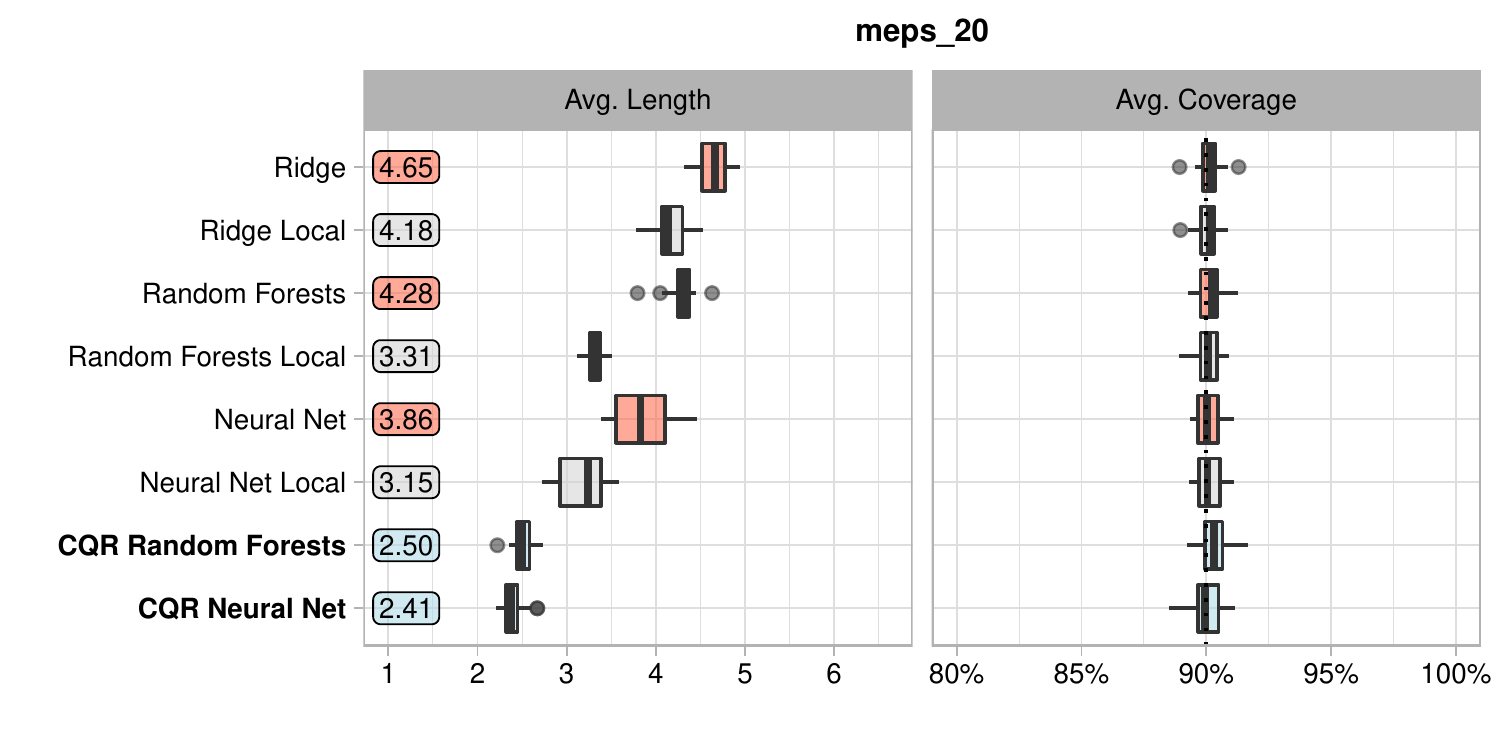}	\includegraphics[width=0.95\textwidth]{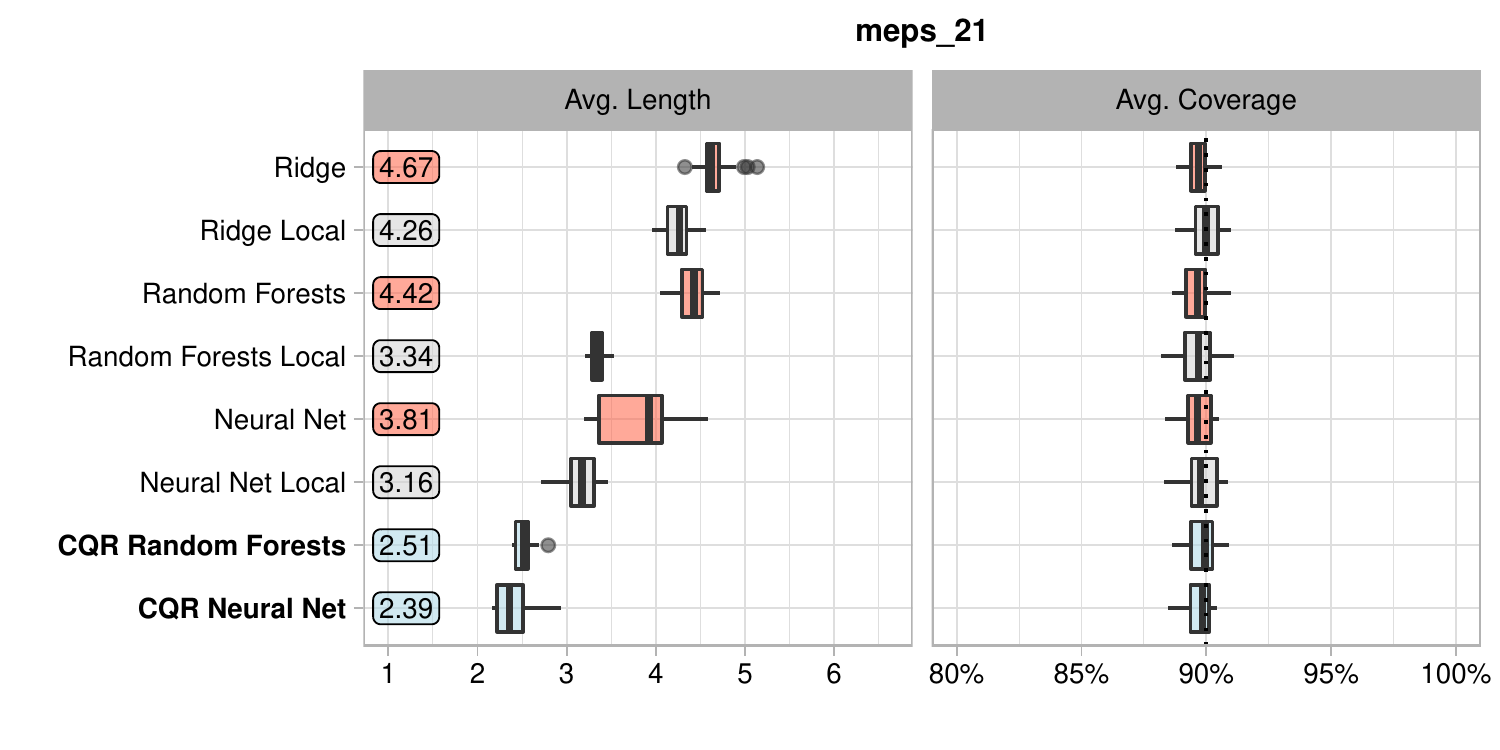}
	\caption{Average length (left) and coverage (right) of prediction intervals ($\alpha = 0.1$), averaged over 20 random (80\%/20\%) training/test splits. The numbers in the colored boxes are the average lengths, shown in red for split conformal, in gray for locally adapative split conformal, and in light blue for our method. The name of the dataset is located at the top of each plot.
	} 
	\label{fig:results_database_part1}
\end{figure}

\begin{figure}[th]
	\centering
	\includegraphics[width=0.95\textwidth]{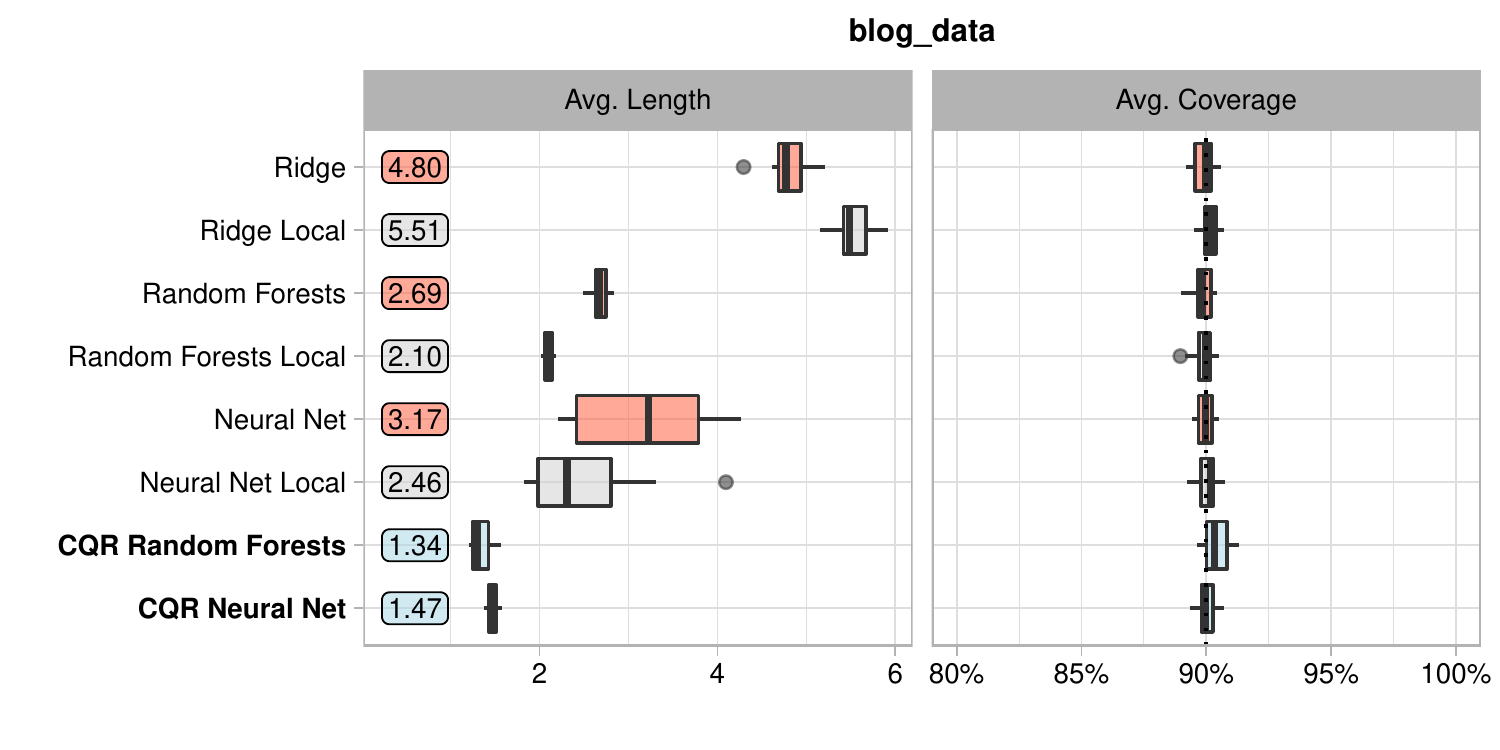}
	\includegraphics[width=0.95\textwidth]{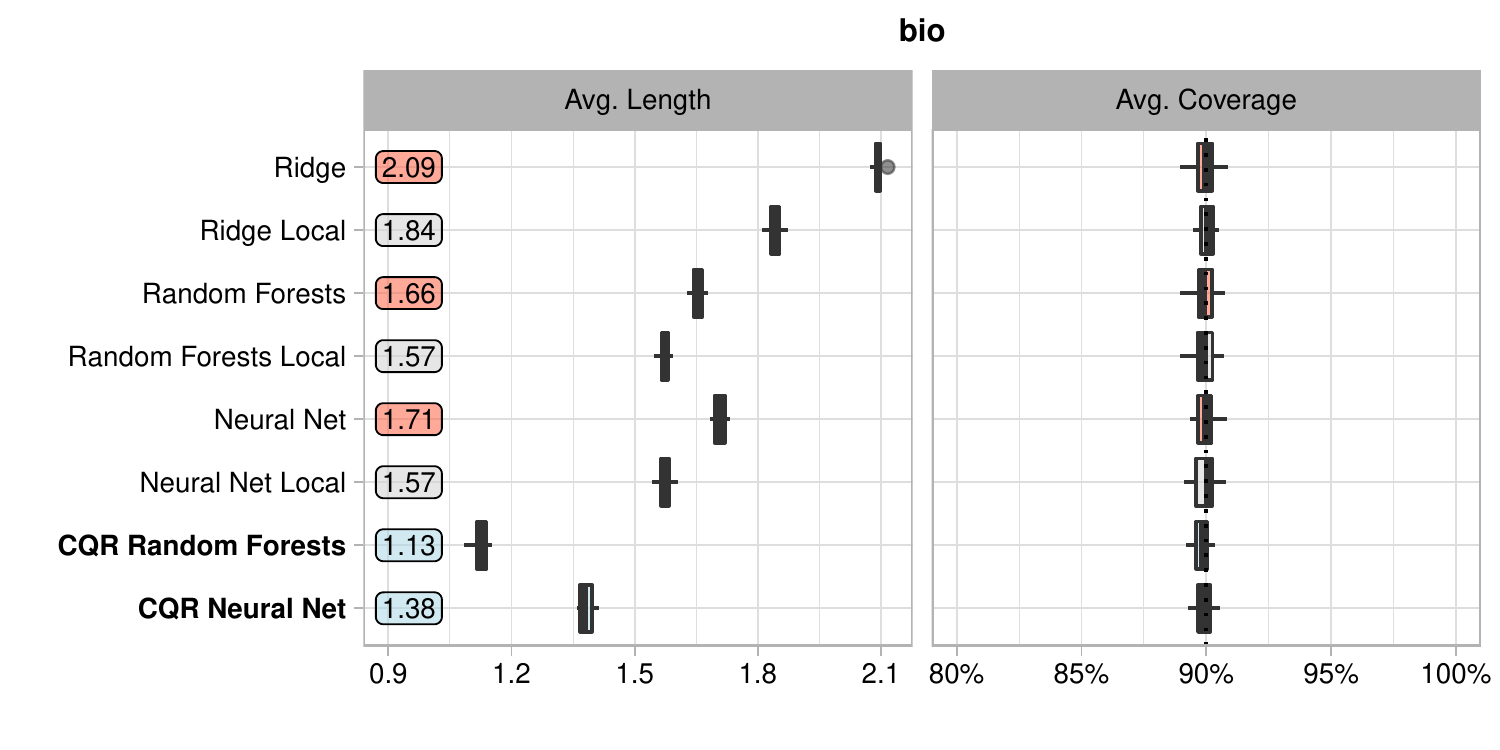}
	\includegraphics[width=0.95\textwidth]{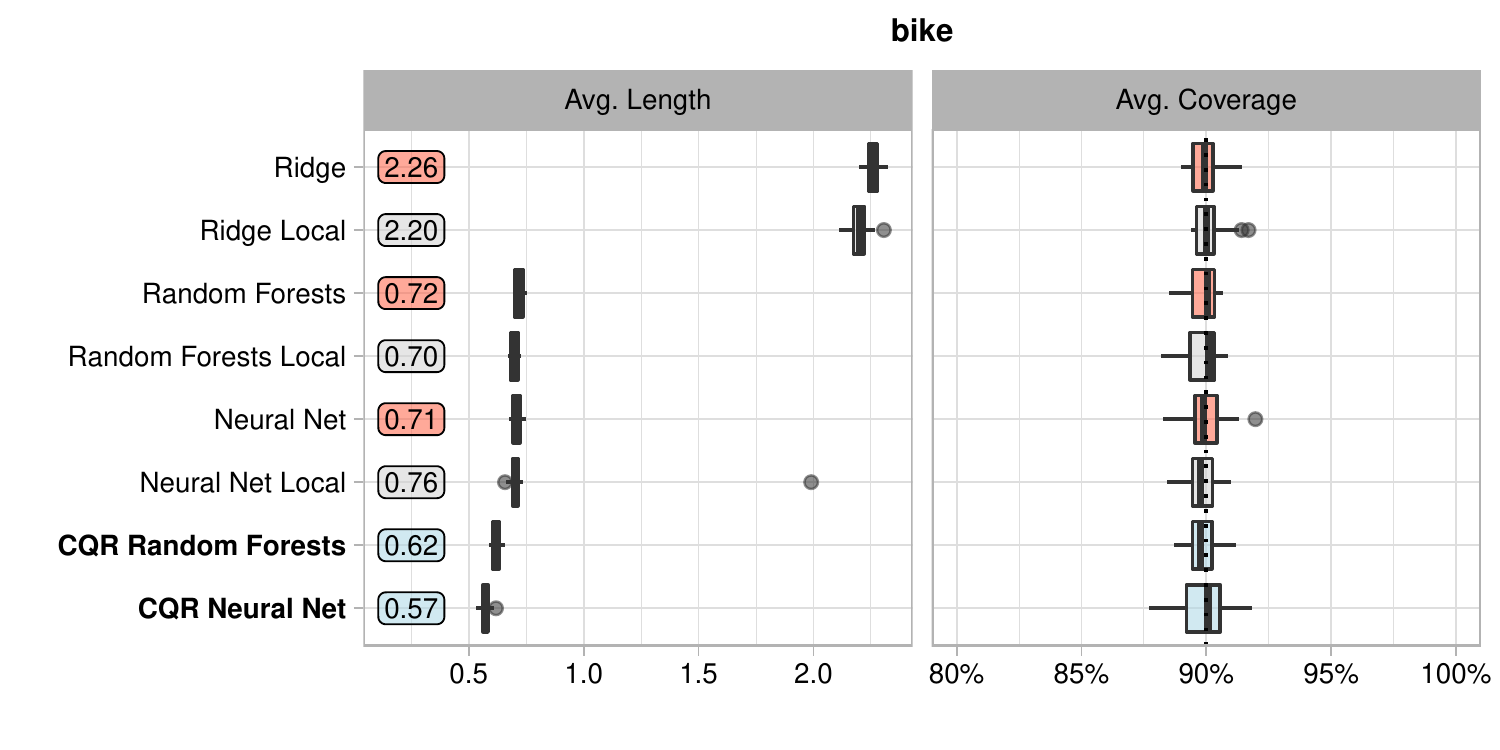}
	\caption{Refer to the caption of Figure \ref{fig:results_database_part1} for details.
	} 
	\label{fig:results_database_part2}
\end{figure}

\begin{figure}[th]
	\centering
	\includegraphics[width=0.95\textwidth]{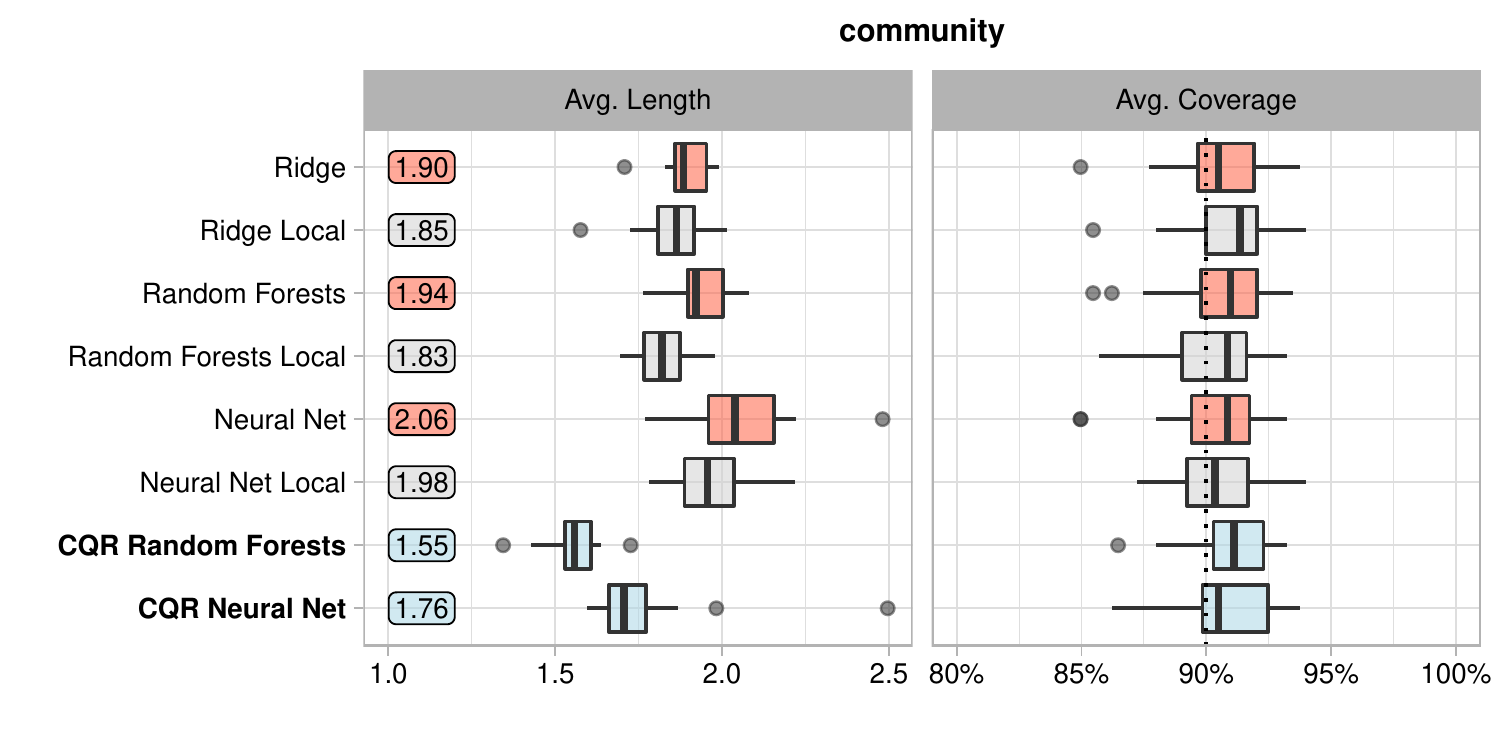}
	\includegraphics[width=0.95\textwidth]{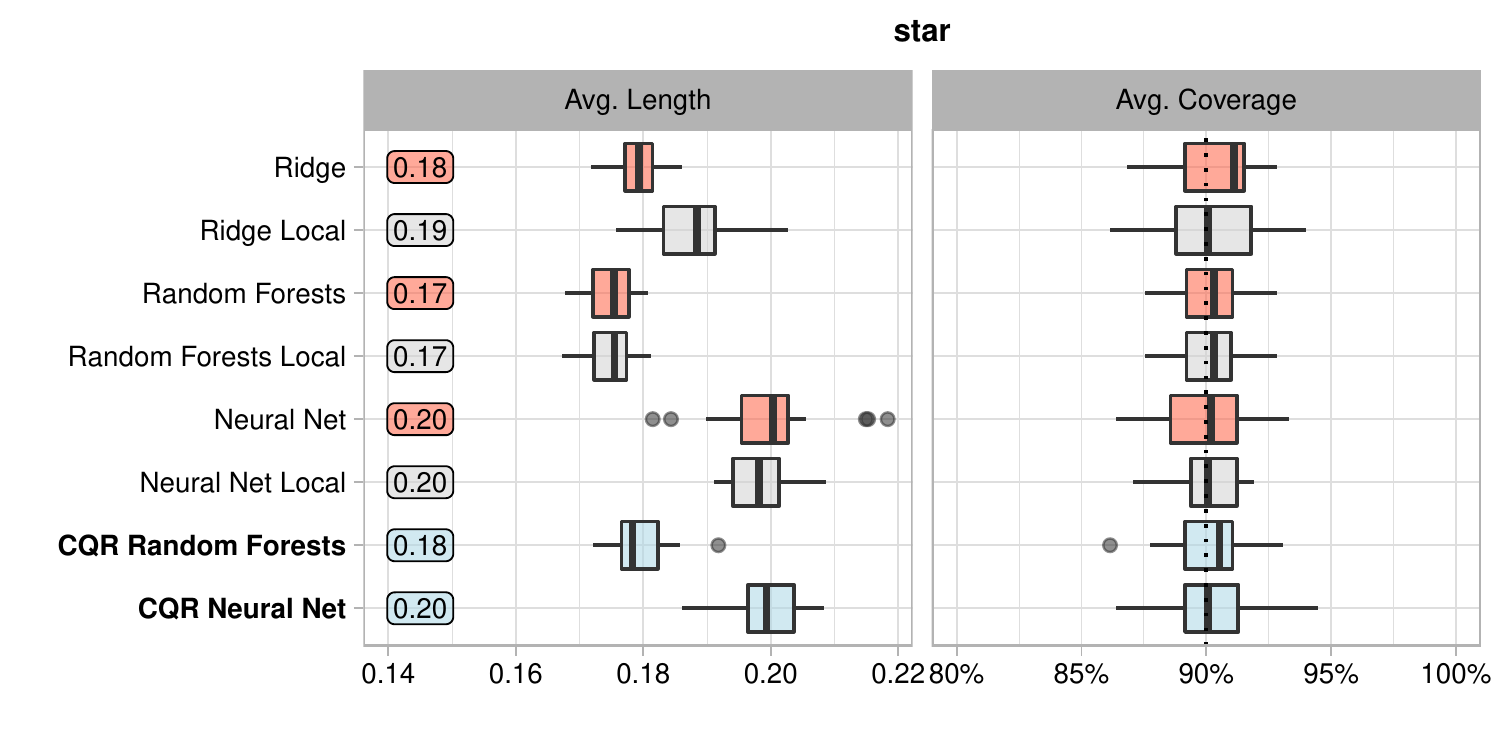}
	\includegraphics[width=0.95\textwidth]{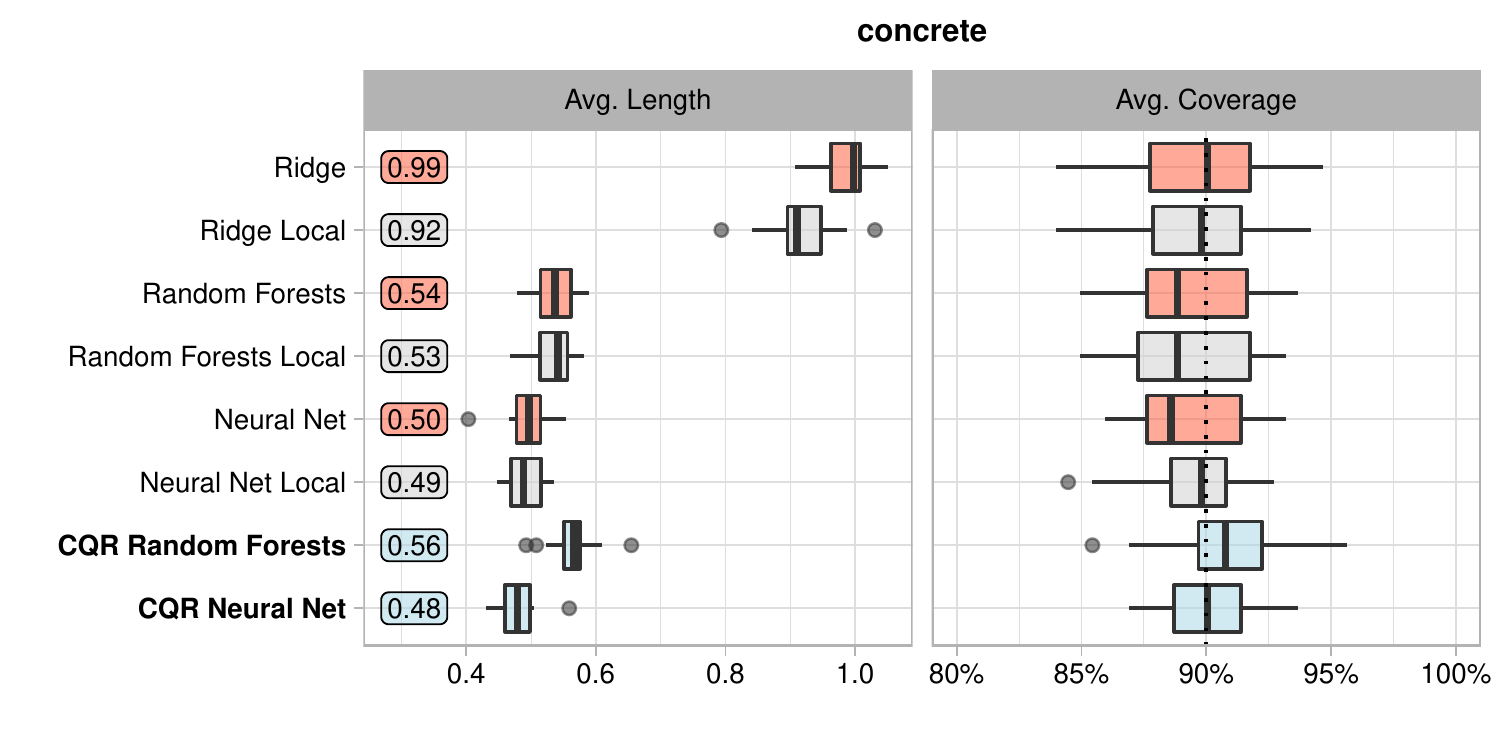}
	\caption{Refer to the caption of Figure \ref{fig:results_database_part1} for details.
	} 
	\label{fig:results_database_part3}
\end{figure}

\begin{figure}[th]
	\centering
	\includegraphics[width=0.95\textwidth]{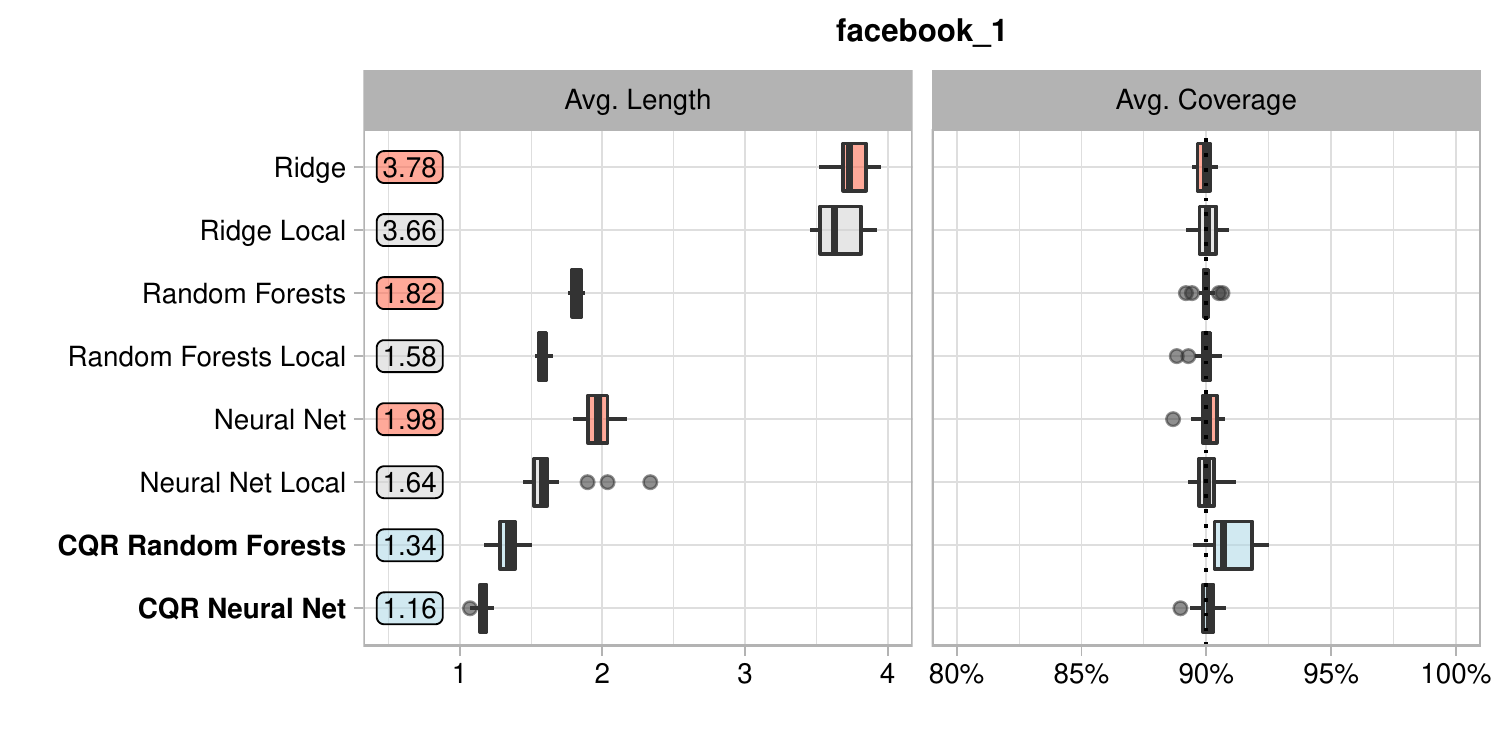}
	\includegraphics[width=0.95\textwidth]{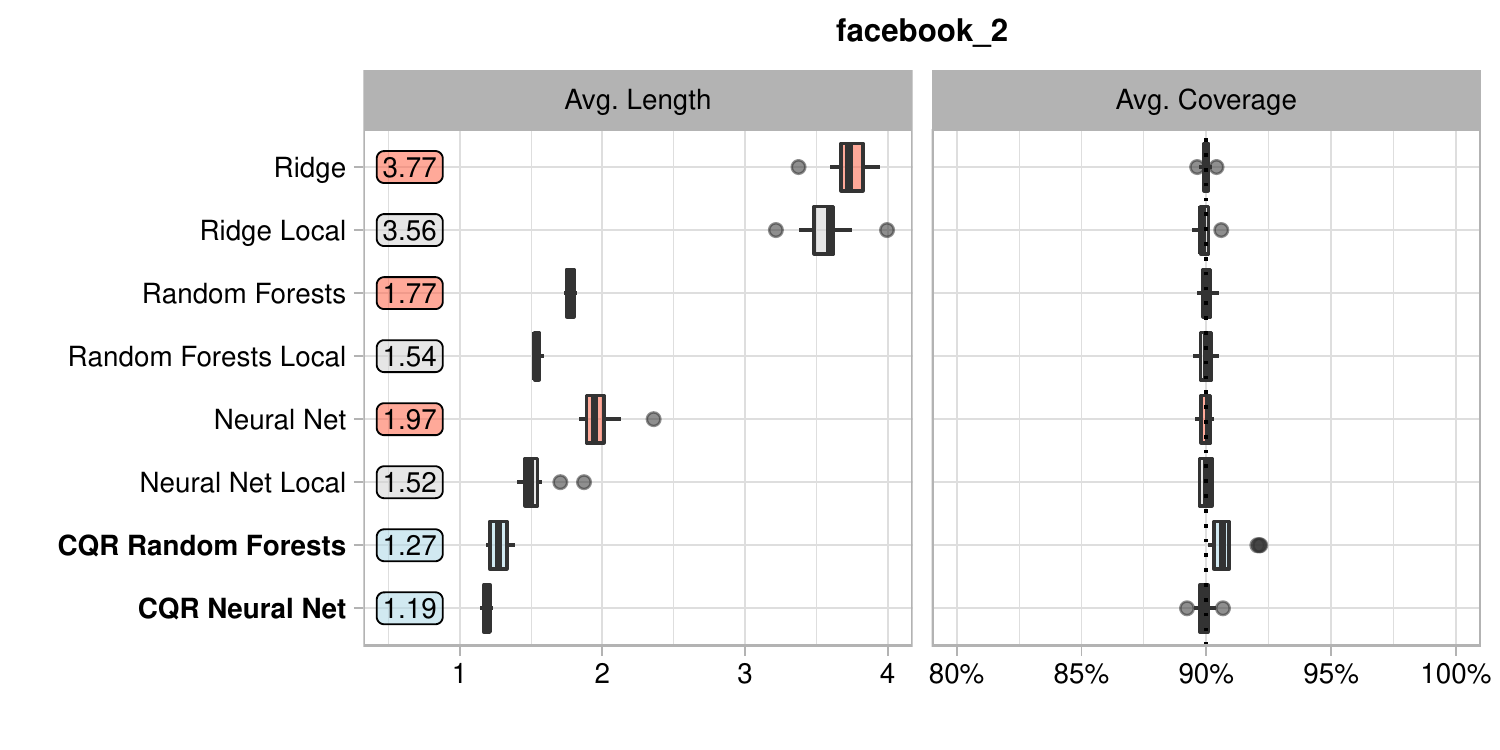}
	\caption{Refer to the caption of Figure \ref{fig:results_database_part1} for details.
	} 
	\label{fig:results_database_part4}
\end{figure}

In a series of figures, we break down the performance of the different methods on each of the benchmark datasets. Figure \ref{fig:results_database_part1} summarizes our experiments on the datasets: medical expenditure panel survey number 19 (\texttt{MEPS\_19}) \cite{meps_19}, number 20 (\texttt{MEPS\_20}) \cite{meps_20}, and number 21 (\texttt{MEPS\_21}) \cite{meps_21}. Figure \ref{fig:results_database_part2} shows the results for: blog feedback (\texttt{blog\_data}) \cite{blog_data}; physicochemical properties of protein tertiary structure (\texttt{bio}) \cite{bio}; and bike sharing (\texttt{bike}) \cite{bike}. Figure \ref{fig:results_database_part3} shows the results for: community and crimes (\texttt{community}) \cite{community}; Tennessee's student teacher achievement ratio (\texttt{STAR}) \cite{star}; and concrete compressive strength (\texttt{concrete}) \cite{concrete}. Lastly, Figure \ref{fig:results_database_part4} shows the results for: Facebook comment volume, variants one (\texttt{facebook\_1}) and two (\texttt{facebook\_2}) \cite{facebook_data,facebook_paper}.

The performance on individual datasets confirms the overall trend in Table \ref{fig:results}. Locally adaptive conformal prediction generally outperforms standard conformal prediction, and, on ten out of eleven datasets, conformalized quantile regression outperforms both. The CQR random forests are overly conservative on the two Facebook datasets. This is consistent with the theory, because in this case there are ties among the conformity scores and so the upper bound in Theorem \ref{thm:validity_qreg} does not apply.

\section{Conclusion} \label{sec:conclusions}
Conformal quantile regression is a new way of constructing prediction intervals that combines the advantages of conformal prediction and quantile regression. It provably controls the miscoverage rate in finite samples, under the mild distributional assumption of exchangeability, while adapting the interval lengths to heteroskedasticity in the data.

We expect the ideas behind conformal quantile regression to be applicable in the related setting of conformal predictive distributions \cite{vovk2017nonparametric}. In this extension of conformal prediction, the aim is to estimate a predictive probability distribution, not just an interval. We see intriguing connections between our work and a very recent, independently written paper on conformal distributions \cite{vovk2019conformal}.

\subsection*{Acknowledgements}
E. C. was partially supported by the Office of Naval Research (ONR)
under grant N00014-16- 1-2712, by the Army Research Office (ARO) under
grant W911NF-17-1-0304, by the Math + X award from the Simons
Foundation and by a generous gift from TwoSigma. E.~P.~and Y.~R.~were partially supported by the
ARO grant. Y.~R.~was also supported by the same Math + X award.  Y.~R.~thanks the Zuckerman
Institute, ISEF Foundation and the Viterbi Fellowship, Technion, for providing additional research support. We thank Chiara Sabatti for her insightful comments on a draft of this paper and Ryan Tibshirani for his crucial remarks on our early experimental findings.

\bibliographystyle{unsrt}
\bibliography{MyBib}

\FloatBarrier

\appendix

\section{Lemmas about quantiles}
\label{app:lemmas}
Recall that the \emph{quantile function} $Q$ of a random variable $Z$, with cumulative distribution function $F(z) := \PP\{Z \leq z\}$, is defined by the equivalence
\begin{equation}
    Q(\alpha) \leq z \quad\text{if and only if }\quad \alpha \leq F(z)
\end{equation}
for all $\alpha \in (0,1)$ and $z \in \RR$. Dually, but less standardly, the \emph{right quantile function} $R$ of the random variable $Z$ is defined by the equivalence
\begin{equation}
    F^-(z) \leq \alpha \quad\text{if and only if}\quad z \leq R(\alpha),
\end{equation}
where $F^-(z) := F(z-) = \PP\{Z < z\}$. The quantile functions have the explicit formulas
\begin{equation}
     Q(\alpha) = \inf\{ z \in \RR: \alpha \leq F(z) \}, \qquad
     R(\alpha) = \sup\{ z \in \RR: F^-(z) \leq \alpha \}.
\end{equation}
As a special case, the \emph{empirical quantile function} $\hat Q_n$ of random variables $Z_1,\dots,Z_n$ is the quantile function with respect to the empirical CDF $\hat F_n(z) := \frac{1}{n} \sum_{i=1}^n 1_{Z_i \leq z}$. Likewise, the \emph{right empirical quantile function} $\hat R_n$ of $Z_1,\dots,Z_n$ is the right quantile function with respect to $\hat F_n^-(z) = \frac{1}{n} \sum_{i=1}^n 1_{Z_i < z}$. They have the explicit formulas
\begin{equation}
    \hat Q_n(\alpha) = Z_{(\ceil{\alpha n})}, \qquad
    \hat R_n(\alpha) = Z_{(\floor{\alpha n}+1)},
\end{equation}
where $Z_{(k)}$ denotes the $k$th smallest value in $Z_1,\dots,Z_n$.

Variants of the following lemmas appear in the literature \cite{vovk2005algorithmic,lei2018distribution,barber2019conformal}. In the interest of clarity and a self-contained exposition, we state and prove them here.

\begin{lemma}[Quantiles and exchangeability] \label{lemma:quantile-exchangeability}
Suppose $Z_1,\dots,Z_n$ are exchangeable random variables. For any $\alpha \in (0,1)$,
\begin{equation}
    \PP\{Z_n \leq \hat Q_n(\alpha)\} \geq \alpha.
\end{equation}
Moreover, if the random variables $Z_1,\dots,Z_n$ are almost surely distinct, then also
\begin{equation}
    \PP\{Z_n \leq \hat Q_n(\alpha)\} \leq \alpha + \frac{1}{n}.
\end{equation}
In this statement, the probabilities are taken over all the variables $Z_1,\dots,Z_n$.
\end{lemma}
\begin{proof}
By exchangeability and the symmetry of $\hat Q_n(\alpha)$ as a function of $Z_1,\dots,Z_n$, the probability $\PP\{Z_i \leq \hat Q_n(\alpha)\}$ is equal to $\PP\{Z_n \leq \hat Q_n(\alpha)\}$ for every $i$. Therefore,
\begin{equation}
\EE\,\hat F_n(\hat Q_n(\alpha))
  = \frac{1}{n} \sum_{i=1}^n \PP\{Z_i \leq \hat Q_n(\alpha)\}
  = \PP\{Z_n \leq \hat Q_n(\alpha)\}.
\end{equation}
By the defining property of the quantile functions, $\hat F_n(\hat Q_n(\alpha)) \geq \alpha$ and $\hat F_n^-(\hat R_n(\alpha)) \leq \alpha$. Moreover, if the samples $Z_1,\dots,Z_n$ are distinct, then $\Vert \hat F_n - \hat F_n^- \Vert_\infty \leq \frac{1}{n}$, and since $\hat Q_n \leq \hat R_n$, we have $\hat F_n(\hat Q_n(\alpha)) \leq \hat F_n(\hat R_n(\alpha)) \leq \hat F_n^-(\hat R_n(\alpha)) + \frac{1}{n} \leq \alpha + \frac{1}{n}$. To complete the proof, take expectations of the inequalities $\hat F_n(\hat Q_n(\alpha)) \geq \alpha$ and $\hat F_n(\hat Q_n(\alpha)) \leq \alpha + \frac{1}{n}$.
\end{proof}

\begin{lemma}[Inflation of quantiles] \label{lemma:quantile-inflation}
Suppose $Z_1,\dots,Z_{n+1}$ are exchangeable random variables. For any $\alpha \in (0,1)$,
\begin{equation}
    \PP\{Z_{n+1} \leq \hat Q_n((1+\tfrac{1}{n})\alpha)\} \geq \alpha.
\end{equation}
Moreover, if the random variables $Z_1,\dots,Z_{n+1}$ are almost surely distinct, then also
\begin{equation}
     \PP\{Z_{n+1} \leq \hat Q_n((1+\tfrac{1}{n})\alpha)\} \leq \alpha + \frac{1}{n}.
\end{equation}
\end{lemma}
\begin{proof}
Let $Z_{(k,m)}$ denote the $k$th smallest value in $Z_1,\dots,Z_m$. Then for any $0 \leq k \leq n$, we have
\begin{equation}
    Z_{n+1} \leq Z_{(k,n)}
    \quad\text{if and only if}\quad
    Z_{n+1} \leq Z_{(k,n+1)}.
\end{equation}
Indeed, if $Z_{n+1} \leq Z_{(k,n)}$, then $Z_{(k,n+1)}$ is the larger of $Z_{(k-1,n)}$ and $Z_{n+1}$; in particular, $Z_{(k,n+1)} \geq Z_{n+1}$. Conversely, if $Z_{n+1} \leq Z_{(k,n+1)}$ then also $Z_{n+1} \leq Z_{(k,n)}$ because $Z_{(k,n+1)} \leq Z_{(k,n)}$.

Thus, since $\hat Q_n((1+\tfrac{1}{n})\alpha) = Z_{(\ceil{\alpha(n+1)},n)}$ and $ \hat Q_{n+1}(\alpha) = Z_{(\ceil{\alpha(n+1)},n+1)}$, we have
\begin{equation}
    Z_{n+1} \leq  \hat Q_n((1+\tfrac{1}{n})\alpha)
    \quad\text{if and only if}\quad
    Z_{n+1} \leq  \hat Q_{n+1}(\alpha)
\end{equation}
and, hence,
\begin{equation}
    \PP\{Z_{n+1} \leq \hat Q_n((1+\tfrac{1}{n})\alpha)\}
    = \PP\{Z_{n+1} \leq \hat Q_{n+1}(\alpha)\}.
\end{equation}
To conclude the proof, apply Lemma \ref{lemma:quantile-exchangeability} with $n$ replaced by $n+1$.
\end{proof}

\section{Synthetic experiment} \label{app:synthetic_example}

\begin{figure}[h]
	\centering
	\includegraphics[width=0.5\textwidth]{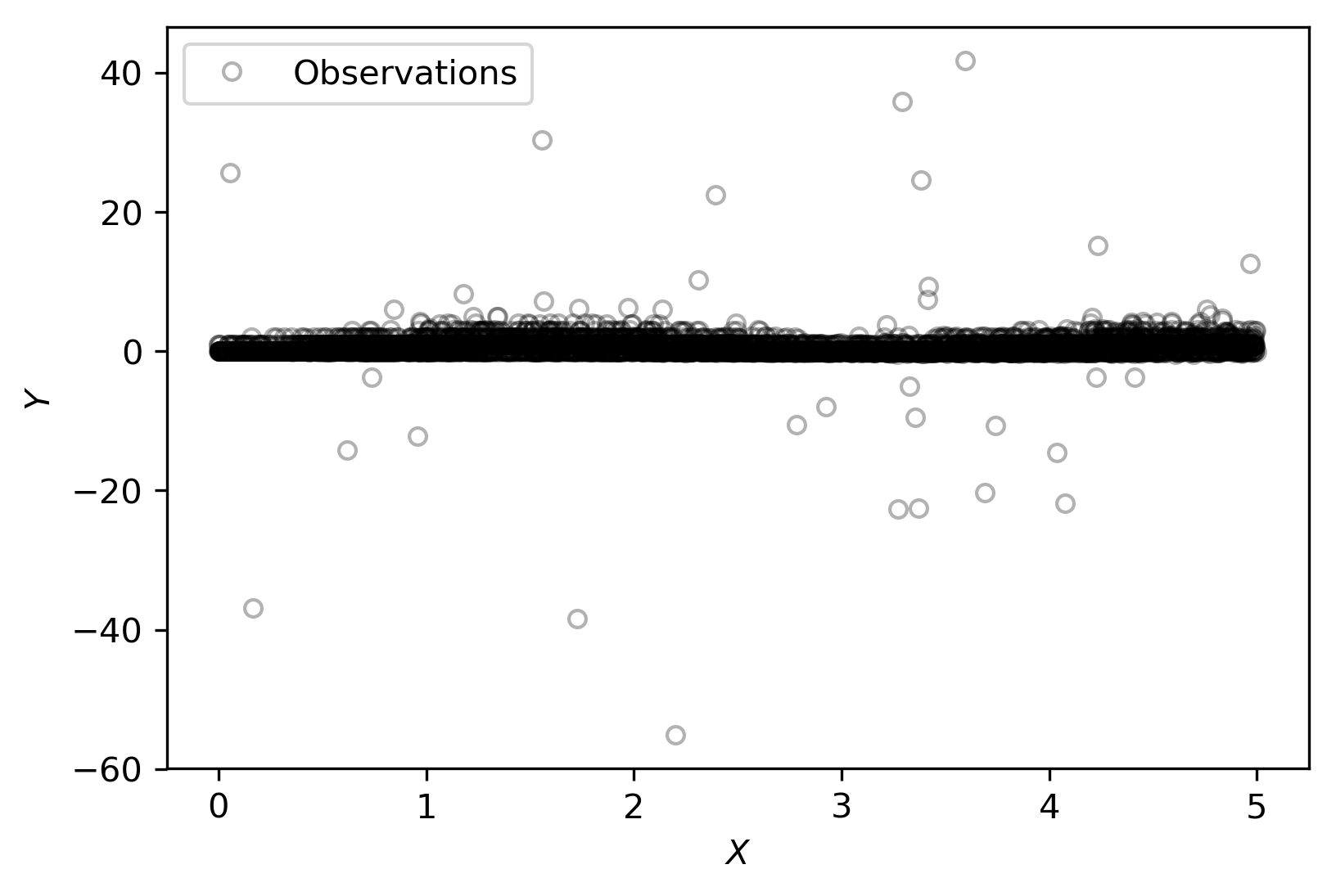}
	\caption{Full range scatter plot of the test data used in the synthetic simulation of Figure \ref{fig:synthetic_illustration}.}
	\label{fig:synthetic_illustration_full}
\end{figure}

In Figure \ref{fig:synthetic_illustration}, we presented an experiment on simulated data to illustrate the importance of adaptivity in conformal prediction. Here we describe the details of that experiment.

To generate the training data, we draw $n=2000$ independent, univariate predictor samples $X_i$ from the uniform distribution on the interval $[1,5]$. The response variable is then sampled as 
\begin{align} \label{eq:synthetic_data}
    Y_i \sim \textrm{Pois}(\sin^2(X_i) + 0.1) + 0.03 \ X_i \ \epsilon_{1,i} + 25 \ \mathbbm{1}\{ U_i < 0.01 \} \ \epsilon_{2,i},
 \end{align}
where $\textrm{Pois}(\lambda)$ is the Poisson distribution with mean $\lambda$, both $\epsilon_{1,i}$ and $\epsilon_{2,i}$ are i.i.d.\ standard Gaussian noise, and the $U_i$'s are uniform on the interval $[0,1]$. We generate a test set of 5000 samples in the same way. The last term in equation \eqref{eq:synthetic_data} creates few but large outliers. This is illustrated in Figure \ref{fig:synthetic_illustration_full}, which, in contrast to Figure \ref{fig:synthetic_illustration}, plots the synthetic data across its full range.

In Figure \ref{subfig:split}, we construct a $90\%$ prediction interval for the test data using split conformal prediction. Specifically, we split the training data into two subsets, train a random forest regressor on the first set, and calibrate the intervals on the second set. In Figure \ref{subfig:local}, we do the same for locally adaptive split conformal prediction. The scale estimator is another random forest. The experiment is insensitive to the value of the hyper-parameter $\gamma$; we set it to zero. Finally, in Figure \ref{subfig:cqr}, we instantiate our method, conformal quantile regression, with quantile random forests \cite{meinshausen2006quantile} as the underlying quantile regression algorithm.

\begin{figure}[ht]
    \centering
	\includegraphics[width=0.5\textwidth]{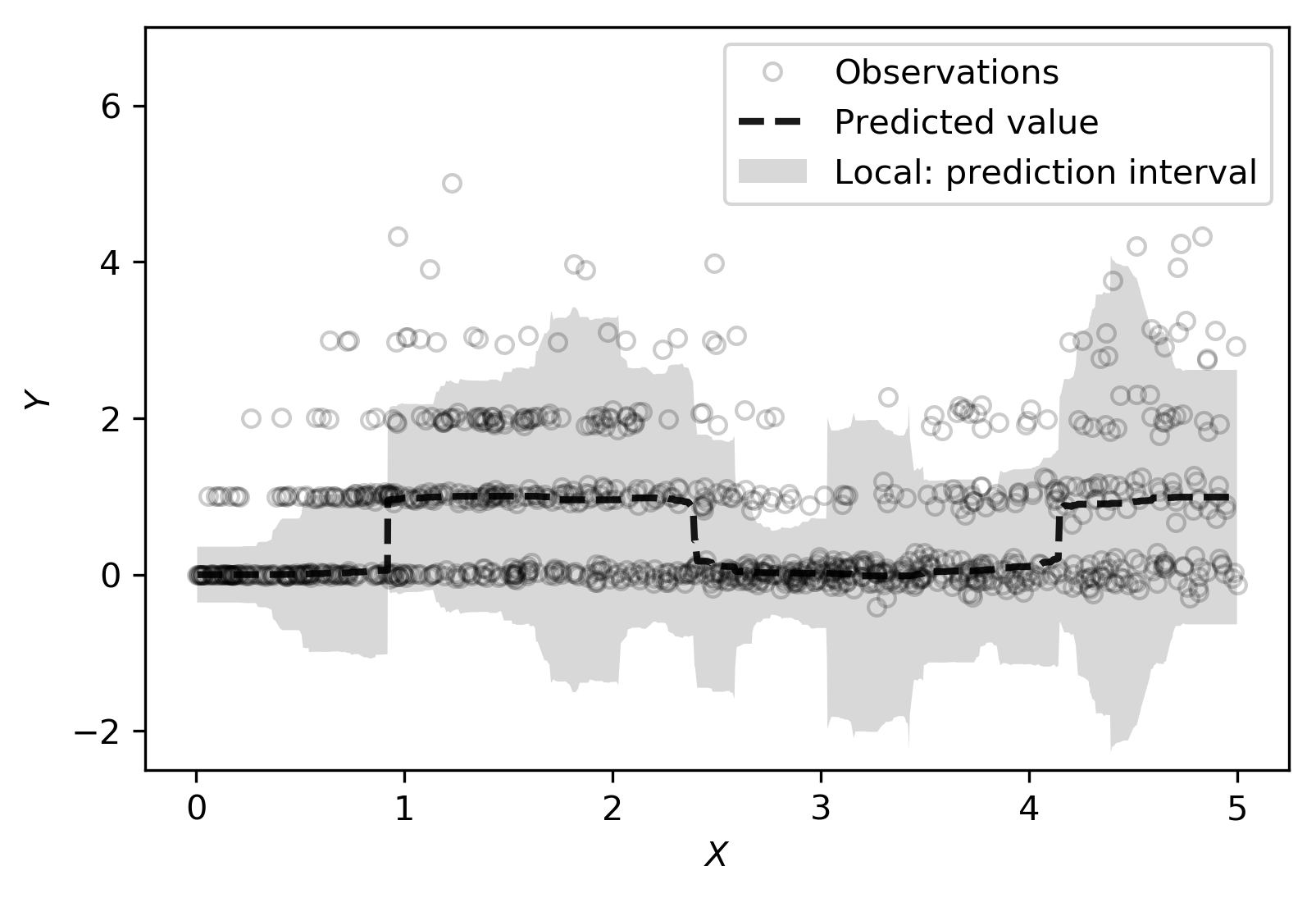}
    \caption{Prediction intervals on simulated data constructed by locally adaptive conformal prediction, with conditional \emph{median} estimation via quantile regression forests. The target coverage is 90\%. On test data, the average coverage is 90.14\% and the average length is 2.86.}
	\label{fig:synthetic_illustration_median}
\end{figure}

To improve robustness to outliers, one might try to estimate the conditional median instead of the conditional mean in locally adaptive conformal prediction. We implement this strategy in Figure \ref{fig:synthetic_illustration_median}, using quantile regression forests \cite{meinshausen2006quantile} to estimate the conditional median. The residuals are scaled in the usual way, by classical regression via random forests. At least on this simulated dataset, estimating conditional medians instead of means has little effect on the average lengths of the prediction intervals (compare Figures \ref{subfig:local} and \ref{fig:synthetic_illustration_median}).

\end{document}